\documentclass[%
 aip,
 amsmath,amssymb,
 reprint,%
groupedaddress
]{revtex4-1}

\usepackage{graphicx}% Include figure files
\usepackage{dcolumn}% Align table columns on decimal point
\usepackage{bm}% bold math
\usepackage[utf8]{inputenc}
\usepackage[T1]{fontenc}
\usepackage[ruled,vlined]{algorithm2e}
\usepackage[mathscr]{euscript}
\DeclareSymbolFont{rsfs}{U}{rsfs}{m}{n}
\DeclareSymbolFontAlphabet{\mathscrsfs}{rsfs}
\DeclareMathOperator*{\argmin}{arg\,min}
\usepackage{xcolor}

\begin{document}

\preprint{AIP/123-QED}

\title[\emph{Efficient hyperparameter tuning for kernel ridge regression with Bayesian optimization}]{Efficient hyperparameter tuning for kernel ridge regression with Bayesian optimization}
% Force line breaks with \\

\author{Annika Stuke}
\affiliation{Department of Applied Physics, Aalto University, P.O. Box 11100, 00076 Aalto, Espoo, Finland}
\author{Patrick Rinke}
\affiliation{Department of Applied Physics, Aalto University, P.O. Box 11100, 00076 Aalto, Espoo, Finland}
\author{Milica Todorovi{\'c}}%
\affiliation{Department of Applied Physics, Aalto University, P.O. Box 11100, 00076 Aalto, Espoo, Finland}

\date{\today}% It is always \today, today,
             %  but any date may be explicitly specified

\begin{abstract}
Machine learning methods usually depend on internal parameters -- so called hyperparameters -- that need to be optimized for best performance. Such optimization poses a burden on machine learning practitioners, requiring expert knowledge, intuition or computationally demanding brute-force parameter searches. We here address the need for more efficient, automated hyperparameter selection with Bayesian optimization. We apply this technique to the kernel ridge regression machine learning method for two different descriptors for the atomic structure of organic molecules, one of which introduces its own set of hyperparameters to the method. We identify optimal hyperparameter configurations and infer entire prediction error landscapes in hyperparameter space, that serve as visual guides for the hyperparameter dependence. We further demonstrate that for an increasing number of hyperparameters, Bayesian optimization becomes significantly more efficient in computational time than an exhaustive grid search -- the current default standard hyperparameter search method -- while delivering an equivalent or even better accuracy.
\end{abstract}

\maketitle

\section{\label{intro}Introduction}

With the advent of datascience \cite{fourthparadigm,Agrawala}, data-driven research is becoming ever more popular in physics, chemistry and materials science \cite{Aykol/etal:2019, Himanen/Geurts/Foster/Rinke:2019, tms, Rampi_review, Zunger:2018}. Concomitantly, the importance of machine learning as a means to infer knowledge and predictions from the collected data is rising. Especially in molecular and materials science, machine learning has gained traction in the last years and now frequently complements other theoretical or experimental methods \cite{Rampi_review,Zunger:2018, ma_deep_2015,shandiz_application_2016, gomez_design_2016, sendek_machine_2018, rlb2018,goldsmith_machine,meyer_machine_2018,Gu/etal:2019,Schmidt/etal:2019,Jensen/etal:2020,Coley/etal:2020}.%\PRc{we could add more ML reviews from 2019 and 2020 here, if we find any} 

The effective use of machine learning usually requires expert knowledge of the underlying model and the problem domain. A particular difficulty that is sometimes overlooked in current machine learning applications is the optimization of internal model parameters, so called hyperparameters. 
%While experienced data scientists usually have a good sense for promising starting points in a given problem domain, 
Non-expert data scientists often spend a long time exploring countless hyperparameter and model configurations for a given dataset before settling on the best one. However, the best settings for these hyperparameters change with different datasets and dataset sizes. The resulting machine learning model is frequently only applicable to one specific problem setting. New data requires hyperparameter re-optimization. Thus, expert use of machine learning is often a costly endeavor -- both in terms of time and computational budget.

Optimal machine learning is achieved with the set of hyperparameters that optimize some score function $f$, such as mean absolute error (MAE), root mean squared error (RMSE), or coefficient of determination ($\mathrm{R}^2$). The score function reflects the quality of learning given the dataset composition and training set size, and is itself an unknown function of all $n$ hyperparameters $f$=$f(\{\boldsymbol{z}\})$. Hyperparameter tuning comprises a set of strategies to navigate the $n$-dimensional phase space of hyperparameters and pinpoint the parameter combination that brings about the best performance of the machine learning model.

The most commonly used form of automated hyperparameter tuning is grid search. Here the hyperparameter search space is discretised and an exhaustive search is launched across this grid. This brute-force technique is widely employed in machine learning libraries: the algorithm is easy to parallelise and is guaranteed to find the best solution. However, the number of possible parameter combinations to be explored grows exponentially with the  number of hyperparameters $n$. Grid search becomes prohibitively expensive if more than 2 or 3 hyperparameters need to be optimized simultaneously. 

One strategy to overcome this problem is to decompose the $n$-dimensional hyperparameter space into $n$ separate 1-dimensional spaces around each parameter, assuming no interdependence between them. The set of 1-dimensional grid searches can then be solved in turn, while keeping the values of other hyperparameters fixed. This approach is rarely pursued since hyperparameters are typically codependent: optimal solutions in each dimension depend on the fixed values and there is no guarantee that the correct overall solution can be found.

%which is an exhaustive brute-force search exploring a pre-defined range and number of model parameters to find the parameter combination that yields the most optimal machine learning model. Grid search has appeal due to its algorithmic simplicity and availability in many machine learning libraries, but the number of possible parameter combinations to be explored grows exponentially with the number of hyperparameters. Due to this averse scaling behavior, grid search becomes prohibitive, if more than 2 or 3 hyperparameters need to be optimized simultaneously.  

Hyperparameter tuning is a classic complex optimization problem, where the objective score function $f(\{\boldsymbol{z}\})$ has an unknown functional form, but can be evaluated at any point. An algorithm designed to address such tasks is Bayesian optimization \cite{srinivas_gaussian}. It is widely applied in the machine learning community to tune hyperparameters of commonly used algorithms, such as random forest, deep neural network, deep forest or kernel methods \cite{wu_hyper_2019, Yogatama2014EfficientTL, Perrone2019LearningSS, olson_evaluation_2016, young_hyperspace_2018}, which are evaluated on a wide range of standard datasets from the UCI machine learning repository \cite{Dua:2019}. However, it is not yet common to apply Bayesian optimization to machine learning problems in the natural sciences, where high-dimensional hyperparameter spaces are frequently encountered. 

%An alternative, more flexible approach to this problem is to view hyperparameter tuning as the optimization of an unknown black-box function and to deploy algorithms developed for such problems. A suitable and efficient choice is Bayesian optimization \cite{srinivas_gaussian}. Bayesian optimization based approaches are widely applied in the machine learning community to tune hyperparameters of commonly used algorithms, such as random forest, deep neural network, deep forest or kernel methods \cite{wu_hyper_2019, Yogatama2014EfficientTL, Perrone2019LearningSS, olson_evaluation_2016, young_hyperspace_2018}, which are evaluated on a wide range of standard datasets from the UCI machine learning repository \cite{Dua:2019}. However, it is not yet common to apply Bayesian optimization to machine learning problems in the natural sciences, where high-dimensional hyperparameter spaces are frequently encountered. 

In this study, we demonstrate the advantages of applying Bayesian optimization to a hyperparameter optimization problem in machine learning of computational chemistry. Our test case is a kernel ridge regression (KRR) machine learning model that maps molecular structures to their molecular orbital energies \cite{stuke_chemical_2019}. We represent the molecular structures with two different descriptors, the Coulomb matrix  \citep{rupp_fast_2012} and the many-body tensor representation \cite{huo_unified_2017}. The KRR method itself requires the optimization of two hyperparameters and one kernel choice. 
The Coulomb matrix is hyperparameter-free, but the many-body tensor representation adds up to 14 more hyperparameters to the model. Through pre-testing, we reduce this number to 2 hyperparameters that affect the model the most.  The largest hyperparameter space we encounter in this approach is therefore 4 dimensional. %Through pre-testing, we reduce this number to 4 hyperparameters that affect the model the most.  The largest hyperparameter space we encounter in this approach is therefore 6 dimensional.

%We compare Bayesian optimization with grid search for a kernel ridge regression (KRR) machine learning model that maps molecular structures to their molecular orbital energies. We represent the molecular structures with two different descriptors, the Coulomb matrix  \citep{rupp_fast_2012} and the many-body tensor representation \cite{huo_unified_2017}. The KRR method itself requires the optimization of 2 hyperparameters and 1 kernel choice. Previous work already gave us the optimal kernels for the Coulomb matrix and the many-body tensor representation \cite{stuke_chemical_2019}. The Coulomb matrix is hyperparameter-free, but the many-body tensor representation adds up to 14 more hyperparameters to the model. Through pre-testing, we reduce this number to 4 hyperparameters that affect the model the most.  The largest hyperparameter space we encounter in this approach is therefore 6 dimensional.

In previous work, we established the optimal kernels for the Coulomb matrix and the many-body tensor representation \cite{stuke_chemical_2019}. We set KRR hyperparameters by means of grid search for three different molecular datasets, after we had optimized the hyperparameters of the molecular descriptors manually beforehand \cite{stuke_chemical_2019}. Here, we take the more rigorous approach and combine the optimization of descriptor and model parameters into a hyperparameter search of up to four dimensions. Higher-dimensional searches are easily feasible with BOSS, but four dimensions already illustrate the efficiency of Bayesian optimization over grid search.

The objective of this manuscript is to evaluate the effectiveness and accuracy of BOSS against grid search in optimization problems with up to four dimensions. We show that already in 4 dimensions, BOSS outperforms grid search in terms of efficiency. %To demonstrate the true potential of BOSS in higher dimensional search scenarios, we include an example of optimizing six hyperparameters simultaneously. 
In addition, we present score function landscapes generated across the hyperparameter phase space, and analyse them as a function of dataset size and composition. These landscapes provide insight into how the behavior of the machine learning model changes across a range of possible model configurations. Such insight helps machine learning practitioners to choose possible starting points for similar optimization problems. 

The manuscript is organized as follows. Section II introduces the basic principle of machine learning with kernel ridge regression and illustrates how molecules are represented to the algorithm. The concept of hyperparameter tuning with grid search and Bayesian optimization is explained. In Section III, these two methods are applied to adjust the hyperparameters for our kernel ridge regression model which predicts molecular energies of three molecular datasets. We visualize and discuss our results. Conclusions and outlook are presented in the last section.

%Even when the optimal hyperparameters and model configurations were found for one problem, the model is not per se applicable to other datasets, and more tuning is required.

%We show that grid search cannot efficiently optimize more than two hyperparameters, while BOSS succeeds in optimizing a four dimensional model fast and accurately. Moreover, we aim to give a practical guideline and to offer results to non-expert data scientists employing machine learning in chemcial physics.

%the surrogate probabilistic model by finding the hyperparameters that perform best on the surrogate.   These hyperparameters are then evaluated on the objective function and the surrogate model is refined.

\section{methods}
\subsection{Machine learning model}

We employ kernel ridge regression (KRR) to predict molecular orbital energies of three distinct datasets of organic molecules: the QM9 dataset of 134k small organic molecules \citep{ramakrishnan_quantum_2014}, the AA dataset of 44k conformers of proteinogenic amino acids \citep{ropo_first_2016} and the OE dataset of 62\,k organic molecules \cite{stuke_atomic_2020}. The three datasets have been described in detail in our previous KRR work and we refer the interested reader to Ref.~\onlinecite{stuke_chemical_2019} or the original references for each dataset for more information.

%\PRc{I change the symbol for the molecular representation. You had used $\boldsymbol{x}$, but that is then also used in the MBTR for the x-axis and again in the hyperparameter tuning section. So I change $\boldsymbol{x}$ here to $\boldsymbol{M}$}
In KRR, a scalar target property, here the energy of the highest occupied molecular orbital (HOMO), is expressed as a linear combination of kernel functions $k(\boldsymbol{M},\boldsymbol{M'})$
\begin{equation}
E^{\textnormal{pred}}(\boldsymbol{M})=\sum^{N}_{i=1} w_i k(\boldsymbol{M},\boldsymbol{M}_i).
 \label{eq:predict}
\end{equation}
$\boldsymbol{M}_i$ is the descriptor for molecule $i$ and the sum runs over all training molecules. $w_i$ are the regression weights that need to be learned. 

In the scope of this work, we employ two kernel functions: the Gaussian kernel
\begin{equation}
k_{\textnormal{Gaussian}}(\boldsymbol{M},\boldsymbol{M}')=e^{-\frac{||{\boldsymbol{M}-\boldsymbol{M}'}||_2^2}{2\gamma^2}},
    \label{eq:gaussian}
\end{equation}
which is a function of the Euclidean distance between two molecules $\boldsymbol{M}$, $\boldsymbol{M}'$, and the Laplacian kernel 
\begin{equation}
    k_{\textnormal{Laplacian}}(\boldsymbol{M},\boldsymbol{M}')=e^{-\frac{||{\boldsymbol{M}-\boldsymbol{M}'}||_1}{\gamma}},
    \label{eq:laplacian}
\end{equation}
which is based on the 1-norm to measure similarity between two molecules. In both cases, $\gamma$ is the kernel width that determines the resolution in molecular space.

The regression parameters $w_i$ are obtained from the minimization problem
\begin{equation}
 \underset{w}{\textnormal{min}} \sum_{i=1}^N (E^{\textnormal{pred}} (\boldsymbol{M}_i) - E^{\textnormal{ref}}_i)^2 + \alpha \boldsymbol{w}^T \mathbf{K} \boldsymbol{w},
 \label{eq:krr_minimization}
\end{equation}
where $E^{\textnormal{ref}}_i$ are the known reference HOMO energies in the dataset, $\mathbf{K}$ is the kernel matrix ($K_{i,j}:=k(\boldsymbol{M}_i, \boldsymbol{M}_j$)) and $\boldsymbol{w}$ is the regression weight ($w_i$) vector. The scalar $\alpha$ controls the size of a regularization term and penalizes complex models with large regression weights over simpler models with small regression weights. Equation~\ref{eq:krr_minimization} has an analytic solution 
\begin{equation}
\boldsymbol{w} = (\mathbf{K} + \alpha \mathbf{I})^{-1} \mathbf{E}^{\textnormal{ref}}
\label{eq:solution}
\end{equation}
that determines $\boldsymbol{w}$.

We here explicitly distinguish between the regression weights $w_i$ and the hyperparameters of the machine learning model. The regression weights grow in number with increasing training data size and are given in closed mathematical form by eq.~\ref{eq:solution}. Conversely, the hyperparameters are finite in number.  In KRR, for example, the number of hyperparameters is fixed to two: $\alpha$ and $\gamma$. These two hyperparameters can assume any value within certain sensible ranges. Their optimal values have to be determined by an external hyperparameter tuning procedure. In addition, there are model-specific choices, which could be interpreted as special hyperparameters, that can only assume certain values. For KRR, this would be the choice of kernel. 

%. Both are \emph{learned} during training. However, one class is usually of finite size, whereas the other grows in size with increasing data. In KRR, the regression weights $w_i$ are the model parameter. There is one regression weight per molecule in the training set and the number of model parameters therefore grows with the size of the training set. The hyperparameters in KRR are given by $\gamma$ and $\alpha$. The number of KRR hyperparameters is thus fixed to two and does not grow with dataset size, although the values of $\gamma$ and $\alpha$ may depend on dataset size. In addition, there are model-specific choices, which could be interpreted as special hyperparameters that can only assume certain values. For KRR, this would be the choice of kernel. Here we try two different kernels, the Gaussian and the Laplacian, so the associated hyperparameter would assume integer values of 0 (Gaussian) and 1 (Laplacian).

\subsection{\label{sec:representation}Molecular representation}
One important aspect in machine learning is the representation of the input data to the machine learning algorithm. Here we employ the Coulomb matrix (CM) \citep{rupp_fast_2012} and the many-body tensor representation (MBTR) \cite{huo_unified_2017}. We use the \texttt{DScribe} package \citep{dscribe} to generate both descriptors for the datasets in this work. 

The entries of the CM are given by
\begin{equation}
  C_{ij} = 
  \begin{cases}
    0.5Z_i^{2.4} & \text{if $i=j$}\\
    \frac{Z_i Z_j}{\lVert \mathbf{R_i}-\mathbf{R_j} \rVert } & \text{if $i \neq j$}\\
  \end{cases} .
\end{equation}
The CM encodes the nuclear charges $Z_i$ and corresponding Cartesian coordinates $R_i$ of all atoms $i$ in molecule $\boldsymbol{M}$. The off-diagonal elements represent a \emph{Coulomb repulsion} between atom pairs and the diagonal elements have been fitted to the total energy of the corresponding atomic species in the gas phase. To enforce permutational invariance, the rows and columns of the CM are sorted with respect to their $\ell^2$-norm.

\begin{figure}[h!]
    \centerline{\includegraphics[width=0.9\columnwidth]{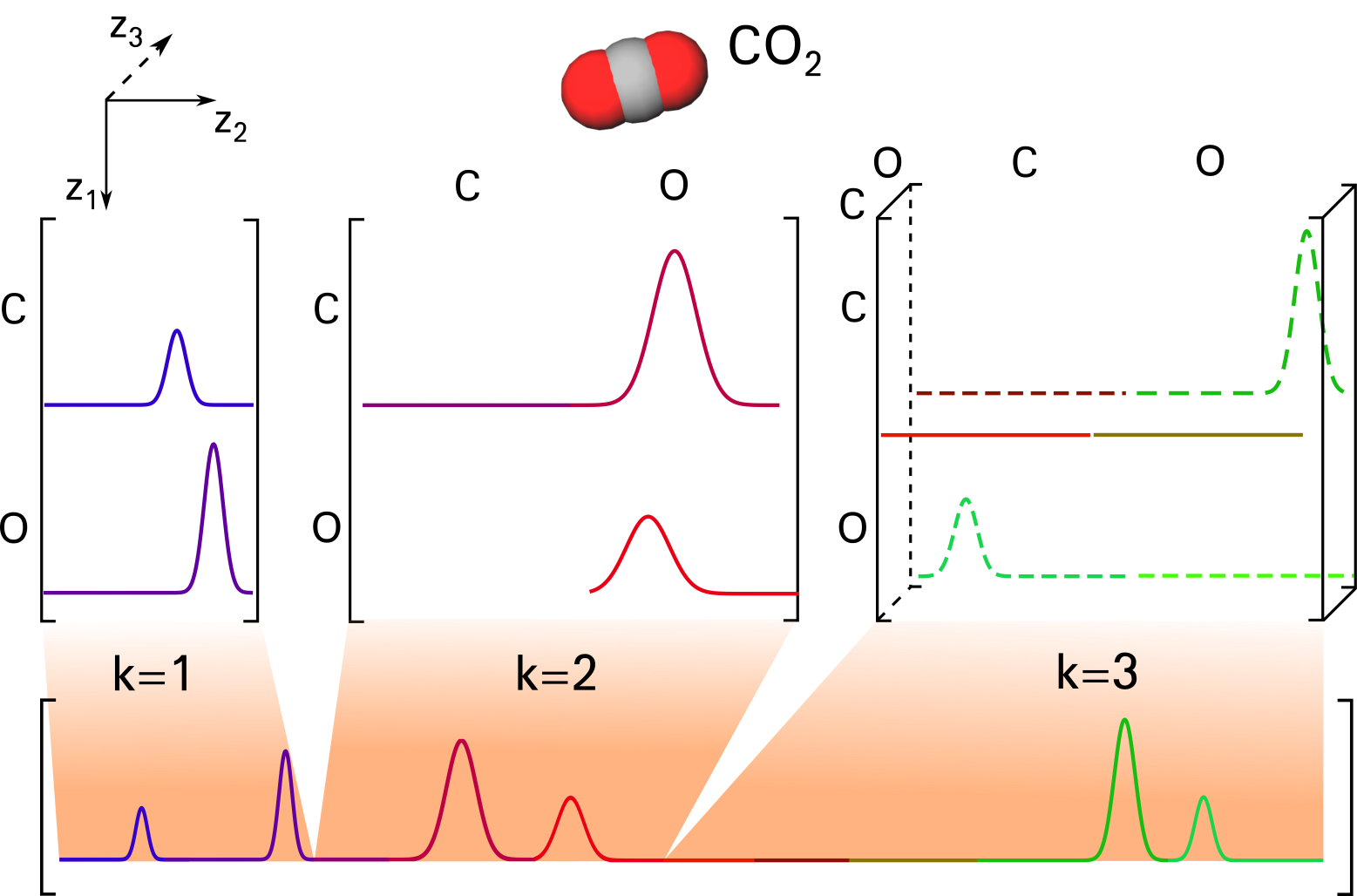}}
    \caption{Illustration of the MBTR output for a CO$_2$ molecule, showing the distributions MBTR$_k$ for $k$=1,2,3 with different combinations of chemical elements. The distributions of each $k$ term are arranged into a $k+1$ dimensional tensor.}
\label{fig:mbtr}
\end{figure}

The MBTR encodes molecular structures by decomposing them into a set of many-body terms (species, interatomic distances, bond angles, dihedral angles, etc.), as outlined for the example of a CO$_2$ molecule in Figure \ref{fig:mbtr}. Each many-body level is represented by a set of fixed sized vectors. The symbol $k$ enumerates the many-body level. We here include terms up to $k$=3. One-body terms ($k$=1) encode all atom types (species) present in the molecule. Two-body terms ($k$=2) encode pairwise inverse distances between any two atoms (bonded and non-bonded). Three-body terms ($k$=3) add angular distributions for any triple of atoms. A geometry function $g_k$ is used to transform each configuration of $k$ atoms into a single scalar value. These scalar values are then Gaussian broadened into continuous representations $\mathcal{D}_k$:

\begin{equation}
\mathcal{D}_{1}^{l}(x)=\frac{1}{\sigma_1 \sqrt{2 \pi}}e^{-\frac{(x-g_1(Z_l))^2}{2\sigma_1^2}}
\end{equation}

\begin{equation}
\mathcal{D}_{2}^{l,m}(x)=\frac{1}{\sigma_2 \sqrt{2 \pi}}e^{-\frac{(x-g_2(\boldsymbol{R}_l, \boldsymbol{R}_m))^2}{2\sigma_2^2}}
\end{equation}
\begin{equation}
\mathcal{D}_{3}^{l,m,n}(x)=\frac{1}{\sigma_3 \sqrt{2 \pi}}e^{-\frac{(x-g_3(\boldsymbol{R}_l, \boldsymbol{R}_m), \boldsymbol{R}_n)^2}{2\sigma_3^2}}.
\end{equation}
The $\sigma_k$'s are the feature widths for the different $k$-levels and $x$ runs over a predefined range $[x_\textrm{min}^k,x_\textrm{max}^k]$ of possible values for the geometry functions $g_k$. For $k=1, 2, 3$, the geometry functions are given by $g_1(Z_{l})=Z_{l}$ (atomic number), $g_2(\boldsymbol{R}_{l}, \boldsymbol{R}_{m})=|\boldsymbol{R}_{l}- \boldsymbol{R}_{m}|$ (distance) or $g_2(\boldsymbol{R}_{l}, \boldsymbol{R}_{m})=\frac{1}{|\boldsymbol{R}_{l}- \boldsymbol{R}_{m}|}$ (inverse distance), and $g_3(\boldsymbol{R}_{l}, \boldsymbol{R}_{m}, \boldsymbol{R}_{n})=\textnormal{cos}(\angle(\boldsymbol{R}_{l}- \boldsymbol{R}_{m}, \boldsymbol{R}_{n}- \boldsymbol{R}_{m}))$ (cosine of angle). For each possible combination of chemical elements present in the dataset, a weighted sum of distributions $\mathcal{D}_k$ is generated. For $k=1,2,3$, these final distributions are given by
\begin{equation}
\textnormal{MBTR}_{1}^{Z_1}(x) = \sum_{l}^{|Z_1|}w_1^l\mathcal{D}
_1^l(x)
\end{equation}

\begin{equation}
\textnormal{MBTR}_{2}^{Z_1, Z_2}(x) = \sum_{l}^{|Z_1|} \sum_{m}^{|Z_2|}w_2^{l,m}\mathcal{D}
_2^{l,m}(x)
\end{equation}
\begin{equation}
\textnormal{MBTR}_{3}^{Z_1, Z_2, Z_3}(x) = \sum_{l}^{|Z_1|} \sum_{m}^{|Z_2|} \sum_{n}^{|Z_3|}w_3^{l,m,n}\mathcal{D}
_3^{l,m,n}(x),
\end{equation}
where the sums for $l$, $m$, and $n$ run over all atoms with atomic numbers $Z_1$, $Z_2$ and $Z_3$. $w_k$ are weighting functions that balance the relative importance of different $k$-terms and/or limit the range of inter-atomic interactions. For $k=1$, usually no weighting is used ($w_1^l=1$). For $k=2$ and $k=3$ the following exponential decay functions are implemented in \texttt{DScribe}
\begin{equation}
w_2^{l,m} = e^{-s_k|\boldsymbol{R}_{l}- \boldsymbol{R}_{m}|}
\end{equation}
\begin{equation}
w_3^{l,m,n} = e^{-s_k(|\boldsymbol{R}_{l}- \boldsymbol{R}_{m}|+|\boldsymbol{R}_{m}- \boldsymbol{R}_{n}|+|\boldsymbol{R}_{l}- \boldsymbol{R}_{n}|)}
\end{equation}
The parameter $s_k$ effectively tunes the cutoff distance. The functions MBTR$_k(x)$ are then discretized with $n_k$ many points in the respective intervals $[x_\textrm{min}^k,x_\textrm{max}^k]$.

\subsection{Number and choice of hyperparameters}

\begin{table}[ht!]
\centering
 \begin{tabular}{|l|l|c|} 
 \hline
 \multicolumn{1}{|c|}{} & 
 \multicolumn{1}{c|}{Type} & 
 \multicolumn{1}{c|}{Number} \\ \hline \hline
 KRR  & feature width ($\gamma$) & 1 \\ 
      & regularization ($\alpha$) & 1 \\
      & kernel type & 1\\ \hline
 CM  & none & 0\\ \hline
 MBTR & $k$-term feature widths ($\sigma_k$) & 3 \\
      & weighting factors ($s_k$) & 2 \\
      & discretization ($[x_\textrm{min}^k,x_\textrm{max}^k]$, $n_k$) & 9 \\ \hline
 \end{tabular}
\caption{List of hyperparameter types and their total number in KRR, the CM and the MBTR.}
 \label{table:HPs} 
\end{table}

In this section we review the hyperparameter types in our CM- or MBTR-based KRR models and motivate our choice for which hyperparameters to investigate in more detail. Table \ref{table:HPs} gives an overview over all hyperparameters in this work. In total there would be 3 hyperparameters to optimize for CM-KRR and 17 for MBTR-KRR. Previous work has shown that some hyperparameters have little effect on the model. They can be preoptimized and set as defaults for the optimization of the remaining hyperparameters. We will explain this choice in more detail in the following.

The KRR method has 3 hyperparameters. $\gamma$ and $\alpha$ are continuous variables and need to be optimized. Conversely, the kernel choice can only assume certain finite values (0 and 1 in our case). We found in previous work \cite{stuke_chemical_2019} that the Laplacian kernel is more accurate for the CM representation and the Gaussian kernel for the MBTR. We therefore fix this choice also in this work and only optimize the 2 parameters $\gamma$ and $\alpha$.

The CM has no hyperparameters. Conversely, the MBTR introduces many. This indicates that the MBTR offers a more complex representation that could lead to faster learning for the same machine learning algorithm. This is indeed what we observed in our previous work comparing CM-KRR and MBTR-KRR \cite{stuke_chemical_2019}. However, the learning improvement comes at the price of a large number of hyperparameters, that need to be optimized to achieve a good model.

As Tab.~\ref{table:HPs} illustrates, MBTR introduces a total of 14 hyperparameters. In previous work \cite{stuke_chemical_2019, dscribe}, we found that our KRR models were not sensitive to the grid discretization parameters. We therefore fix the grids to a range [0, 1] for $k=2$ and [-1, 1] for $k=3$, with 200 discretization points, and leave them unchanged for the rest of this study. We also found that the $k$=1 term does not improve the learning for the three datasets under investigation here \cite{stuke_chemical_2019} and therefore omit the $\sigma_1$ hyperparameter. 

The molecules in our datasets are relatively small. We therefore do not need to limit the range of the MBTR and set $s_2=s_3=0$. This leaves us with 2 hyperparameters, the two feature widths $\sigma_2$ and $\sigma_3$. The minimum and maximum values of all hyperparameters in this study are listed in Table \ref{table:search_space}. 

%In this work, we first investigate the optimization of the two MBTR broadening widths $\sigma_2$ and $\sigma_3$, while the MBTR weighting parameters $s_{2}$ and $s_{3}$ are held constant at their default value of 0.5. Then we also include the weighting parameters and optimize all 6 hyperparamers jointly. The minimum and maximum values of all hyperparameters in this study are listed in Table \ref{table:search_space}. 

\begin{table}
\centering
 \begin{tabular}{|c|c|c|}
 \hline
 Hyperparameter & lower bound & upper bound \\
 \hline
 \hline
 $\alpha$ & 1e-10  & 1  \\ 
 \hline
 $\gamma$ & 1e-10  & 1e-3  \\
 \hline
 $\sigma_2$ & 1e-6 & 1 \\
 \hline
 $\sigma_3$ & 1e-6 & 1 \\
 \hline
 %$s_{2}$ & 0 & 5 \\
 %\hline
 %$s_{3}$ & 0 & 5 \\
 %\hline
 \end{tabular}
 \caption{Hyperparameter search space for BOSS and grid search. }
 \label{table:search_space}
\end{table}

\subsection{Hyperparameter tuning}

Let $\boldsymbol{z}$ be a set of $n$ hyperparameters $\boldsymbol{z}={z_1, z_2, ..., z_n}$, the boundaries of which define the hyperparameter search domain $\mathcal{Z}$, such that $\boldsymbol{z} \in \mathcal{Z}$. The score function $f({\boldsymbol{z}}) \in \mathcal{Z}$ is a \textit{black-box} function defined within the phase space $\mathcal{Z}$. The aim of hyperparameter optimization for a given machine learning model is to find the set of hyperparameters $\boldsymbol{\hat z}$ that provides the best model performance $\hat y$, as measured on a validation set:
\begin{equation}
\boldsymbol{\hat z} =  
\argmin_{\mathbf{z} \in \mathcal{Z}} 
f(\boldsymbol{z}), \hspace{1cm} \hat y=f(\boldsymbol{\hat z})
\end{equation}

%In this study, the score function $f(\boldsymbol{z})$ is the mean absolute error (MAE) on the prediction of HOMO energies, with units in eV.
%where $\boldsymbol{z}$ can take on any value in the domain $\boldsymbol{Z}$. 
The search for $\boldsymbol{\hat z}$ requires sampling the phase space $\mathcal{Z}$ through repeated $f(\boldsymbol{z})$ evaluations. Unfortunately, computing the objective function can be expensive. For each set of hyperparameters, it is necessary to train a model on the training data, make predictions on the validation data, and then calculate the validation metric. With an increasing number of hyperparameters, large datasets and complex models, this process quickly becomes intractable to do by hand. Therefore, automated hyperparameter tuning methods 
%such as grid search and Bayesian optimization 
are indispensable tools for model building in machine learning.

In this study, we compare two approaches for hyperparameter tuning, grid search and Bayesian optimization. The former is guaranteed to find the optimal solution $\boldsymbol{\hat z}$, the latter is a statistical model with a high probability of finding $\boldsymbol{\hat z}$. Our score function $f(\boldsymbol{z})$ is the mean absolute error (MAE) on the prediction of HOMO energies, with units in eV.

\begin{figure}[h!]
    \centerline{\includegraphics[width=0.95\columnwidth]{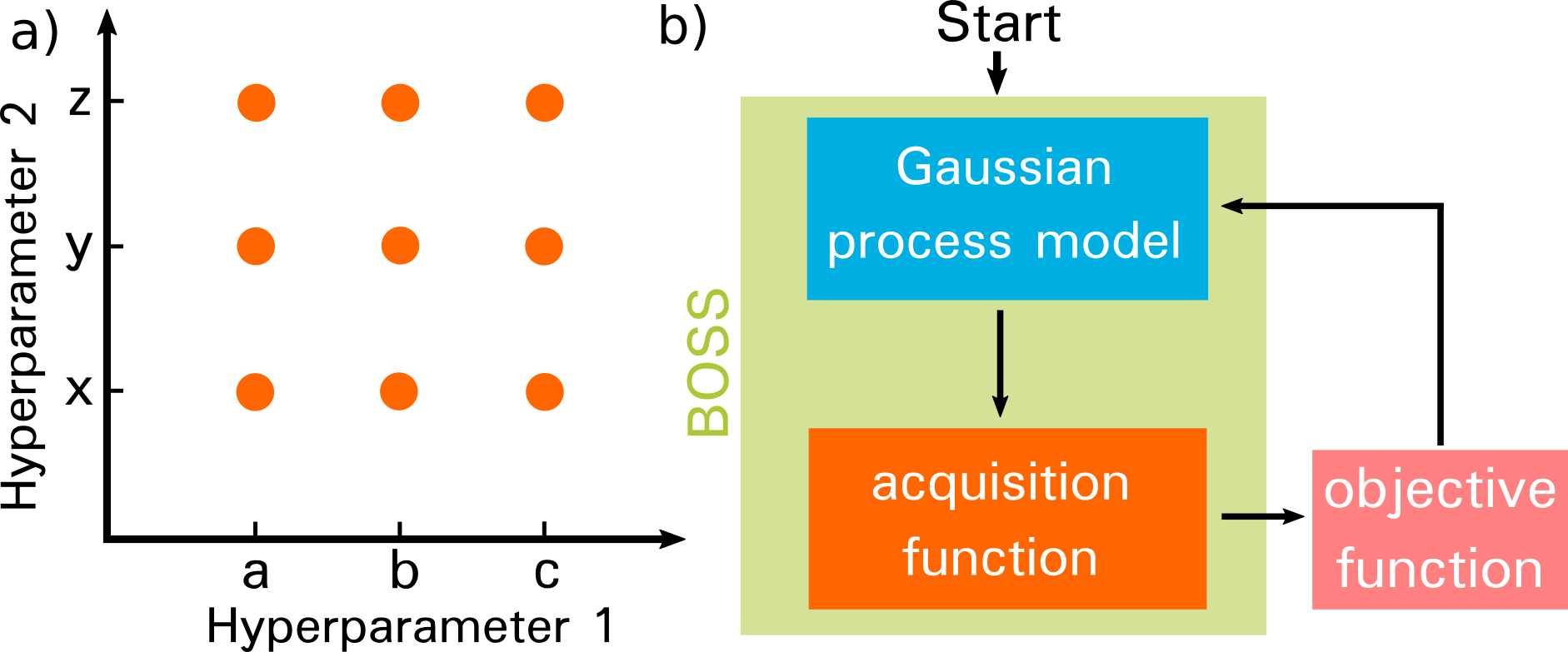}}
    \caption{Working principles of hyperparameter tuning methods. a) In grid search, the hyperparameter space is mapped onto a grid, and the score function is evaluated for each combination of hyperparameter values. b) In Bayesian optimisation, a Gaussian process model is built to simulate the MAE landscape. The landscape is refined by sampling the score function across the hyperparameter space. }
    %The acquisition function assesses the Gaussian process model in order to suggest a new combination of hyperparameters to evaluate on the objective function. The Gaussian process model resembles the objective function and is iteratively updated until convergence is reached.}
\label{fig:gs_boss}
\end{figure}

\subsubsection{Grid Search}
Grid search employs a grid of evenly spaced values for each hyperparameter $z$ to discretise the entire phase space $\mathcal{Z}$, as illustrated in Figure ~\ref{fig:gs_boss} a). The train-predict-evaluate cycle is then run automatically in a loop to evaluate the MAE for all hyperparameter configurations on the grid. Here, we rely on the \textit{scikit-learn} implementation of KRR, but we eschew its native grid search function '\texttt{sklearn.model\_selection.GridSearchCV}' in favour of own algorithms designed speficially for explicit evaluation of computational cost. Algorithms \ref{code_gs_cm} and \ref{code_gs_4d} demonstrate how the 2D and 4D grid searches were performed with different materials descriptors.

%In our KRR problem, the score to minimize is the MAE.

When the CM is used as molecular descriptor, we first compute a CM for each molecule in the dataset, as described in Algorithm \ref{code_gs_cm}. We shuffle the CMs and distribute them into five equally sized groups, along with their corresponding HOMO energies. We then set up a grid for the KRR hyperparameters $\alpha$ and $\gamma$ , with 11 values for each hyperparameter resulting in 121 grid points. Then, a 5-fold cross-validated KRR routine is performed for each possible combination of $\alpha$ and $\gamma$. 

\vspace{0.5cm}
\begin{algorithm}[H]
\SetAlgoLined
compute CM for all molecules \;
shuffle data and split into 5 equally sized groups \;
\For{$\alpha$ in $\{e^j|j=[-10, -9,..., 0]\}$}{
    \For{$\gamma$ in $\{e^j|j=[-10, -9,..., 0]\}$}{
        \For{group$_i$ in [group$_1$, ..., group$_5$]}{
            set group$_i$ as validation set \;
            set 4 remaining groups as training set \;
            train KRR model with $\alpha$, $\gamma$ and Laplacian kernel on training set \;
            validate model on validation set\;
            }
        obtain 5 MAEs and compute average MAE $=\frac{1}{5}\sum_{i=1}^5 \textnormal{MAE}_i$
    }
}
 \caption{Grid search routine for 2D hyperparameter optimization with CM descriptor.}
 \label{code_gs_cm}
\end{algorithm}
\vspace{0.5cm}

As usual in cross-validation, one of the split-off groups is defined as validation set and the remaining four groups combined serve as training set. A KRR model with the Laplacian kernel and the current combination of $\alpha$ and $\gamma$ is trained on the training set and prediction error MAE ($f(\alpha,\gamma)$) is computed on the validation set. 
%We obtain an MAE describing the deviation of the predicted HOMO energies to the reference HOMO energies for the molecules in the validation set. 
For the next round of cross-validation, we define another group as validation set and the remaining four groups as training set to obtain a second MAE. In the same manner, more rounds of cross-validation are performed, until each of the 5 groups has been used as validation set once and as training set 4 times, resulting in 5 MAEs. The average of these serves as an MAE measure of how well the current combination of KRR hyperparameters performs for that grid point. Once we obtain an average MAE for each combination of $\alpha$ and $\gamma$ on the grid, we can pick the combination that results in the lowest MAE. 

For the MBTR, we investigate 2D and 4D hyperparameter optimizations. The 2D case proceeds analogously to the CM KRR hyperparameter search, after we pick values for $\sigma_2$ and $\sigma_3$, and hold them fixed. The algorithm for the 4D case is depicted in Fig.~\ref{code_gs_4d}.  We loop over $\sigma_2$ and $\sigma_3$ on a logarithmic grid of six points each each and build the MBTR for the dataset for those values. Like the CM, this MBTR is then split into 5 subsets for cross validation. We then enter the $\alpha$ and $\gamma$ optimization as we do in the 2D grid search. %A 6D grid search was not attempted in this work, because already the 4D search is prohibitively time consuming. 

%For the MBTR, we investigate 2D and 4D hyperparameter optimizations. The 2D case proceeds analogously to the CM KRR hyperparameter search, after we pick values for $\sigma_2$, $\sigma_3$, $s_2$ and $s_3$ and hold them fixed. The algorithm for the 4D case is depicted in Fig.~\ref{code_gs_4d}. In a 4D search we hold $s_2$ and $s_3$ constant at their default values. We then loop over $\sigma_2$ and $\sigma_3$ on a logarithmic grid of six points each each and build the MBTR for the dataset for those values. Like the CM, this MBTR is then split into 5 subsets for cross validation. We then enter the $\alpha$ and $\gamma$ optimization as we do in the 2D grid search. A 6D grid search was not attempted in this work, because already the 4D search is prohibitively time consuming. 

\vspace{0.5cm}
\begin{algorithm}[H]
\SetAlgoLined
\For{$\sigma_2$ in  $\{e^j|j=[-6, -5,..., 0]\}$}{

    \For{$\sigma_3$ in  $\{e^j|j=[-6, -5,..., 0]\}$}{
    
        compute MBTR for all molecules\;
        shuffle data and split into 5 equally sized groups \;
        
        \For{$\alpha$ in $\{e^j|j=[-10, -9,..., 0]\}$}{
            \For{$\gamma$ in$\{e^j|j=[-10, -9,..., 0]\}$}{
                \For{group$_i$ in [group$_1$, ..., group$_5$]}{
                     set group$_i$ as validation set \;
                     set 4 remaining groups as training set \;
                     train KRR model with $\alpha$, $\gamma$ and Gaussian kernel on training set \;
                     validate model on validation set\;
                     }
                obtain 5 MAEs and compute average MAE $=\frac{1}{5}\sum_{i=1}^5 \textnormal{MAE}_i$
            }
        }
    }    
}
 \caption{Grid search routine for 4D hyperparameter optimization with MBTR descriptor. }
 \label{code_gs_4d}
\end{algorithm}

\subsubsection{Bayesian Optimization}

With Bayesian optimization, we build a \emph{surrogate model} for the MAE across the search domain, then iteratively refine it until convergence \cite{srinivas_gaussian, rasmussen_gaussian, gutmann_bayesian}. Once the MAE surrogate landscape is known, it can be minimized efficiently to find the optimal choice of hyperparameters at its global minimum location. In this work, we rely on the Bayesian Optimization Structure Search (BOSS) package \cite{Todorovic/etal:2019} for simple and robust Bayesian optimization in physics, chemistry and materials science. 

The Bayesian optimization algorithm, illustrated in Figure \ref{fig:gs_boss}b), features a two-step procedure of Gaussian process regression (GPR), followed by an acqusition function. In the GPR, the MAE surrogate model is computed as the posterior mean of a Gaussian process (GP), given MAE data. This step produces an effective landscape of the MAE in hyperparameter space, which can be viewed and analysed. While the posterior mean is the statistically most likely fit to MAE data, the computed posterior variance (uncertainty) indicates which regions of the hyperparameter space are less well known. Both the mean and variance are then used to compute the eLCB acqusition function. The global minimum of the acqusition function points to the combination of hyperparameters in phase space to be tested next. Once this point is evaluated, the resulting MAE is added to the dataset and the cycle repeats. With each additional datapoint, the MAE surrogate model is improved. The method features variance and lengthscale hyperparameters encoded in the radial basis set (RBF) kernel of the Gaussian process, but these are autonomously refined along with the GPR model. 

In this active learning technique, the data is collected at the same time as the model training is perfomed. The acquisition strategy combines data exploitation (searching near known minima) and exploration (searching previously unvisited regions of phase space) to quickly identify important regions of hyperparameter phase space where MAE is low. This allows us to identify the optimal combination of hyperparameters with relatively few MAE evaluations.

BOSS requires only the range of hyperparameters as input, so it can define the phase space domain before it launches a fully automated $n$-dimensional search for the best combination of hyperparameters. Acquisitions are made according to Algorithms \ref{code_boss_2d} and \ref{code_boss_4d}, which serve as the evaluation function. For each new acquisition, the molecular descriptor (either CM or MBTR) is computed and KRR with 5-fold cross validation is performed. The average MAE is returned to BOSS to refine the MAE surrogate model and to perform the next acquisition. 

Once the $n$-dimensional MAE surrogate models are converged, we can evaluate model accuracy qualitatively and model predictions quantitatively. Model predictions are summarised by the location of the global minimum in hyperparameter space $\boldsymbol{\hat z}$, and its value in the surrogate model $\mu(\boldsymbol{\hat z})$. Although $\mu(z)$ values should be close to the true $f(z)$ score function values, we additionally evaluate $f(\boldsymbol{\hat z})$ to validate the match throughout the convergence cycle. This way we can determine that the model does converge, and that it coverges to the true function $f(z)$.

\begin{algorithm}[H]
\SetAlgoLined
%\KwResult{Write here the result }
define boundaries for $\alpha$ and $\gamma$: $\alpha \in \{e^j|j=[-10, 0]\} $, $\gamma \in \{e^j|j=[-10, 0]\}$ \;
define number of iterations: n$_{\textnormal{it}}$=150 \;
Acquisition function suggests new set of $\alpha$, $\gamma$ \;
compute CM for all molecules\;
shuffle data and split into 5 equally sized groups \;
\For{group$_i$ in [group$_1$, ..., group$_5$]}{
    set group$_i$ as validation set \;
    set 4 remaining groups as training set \;
    train KRR model with $\alpha$, $\gamma$ and Laplacian kernel on training set \;
    validate model on validation set\;
    }

obtain 5 MAEs and compute average MAE $=\frac{1}{5}\sum_{i=1}^5 \textnormal{MAE}_i$ \;
return average MAE to GP model

 \caption{2D hyperparameter optimization with CM descriptor: BOSS routine for objective function evaluation.}
 \label{code_boss_2d}
\end{algorithm}

\vspace{0.5cm}

\begin{algorithm}[H]
\SetAlgoLined
%\KwResult{Write here the result }
define boundaries for $\alpha$, $\gamma$,  $\sigma_2$ and $\sigma_3$: $\alpha \in \{e^j|j=[-10, 0]\}$, $\gamma \in\{e^j|j=[-10, 0]\}$, $\sigma2 \in \{e^j|j=[-6, 0]\}$, $\sigma_3 \in \{e^j|j=[-6, 0]\}$ \;
define number of iterations: n$_{\textnormal{it}}$=300 \;
Acquisition function suggests new set of $\alpha$, $\gamma$, $\sigma_2$, $\sigma_3$ \;
compute MBTR for all molecules\;
shuffle data and split into 5 equally sized groups \;
\For{group$_i$ in [group$_1$, ..., group$_5$]}{
    set group$_i$ as validation set \;
    set 4 remaining groups as training set \;
    train KRR model with $\alpha$, $\gamma$ and Gaussian kernel on training set \;
    validate model on validation set\;
}
obtain 5 MAEs and compute average MAE $=\frac{1}{5}\sum_{i=1}^5 \textnormal{MAE}_i$\;
return average MAE to GP model

 \caption{4D hyperparameter optimization with MBTR descriptor: BOSS routine for objective function evaluation.}
 \label{code_boss_4d}
\end{algorithm}

%\subsection{Landscapes in hyperparameter space}
%Not all hyperparameters that are optimized are necessarily equally important. Figure \ref{} shows a qualitative visualization model performance as a function of hyperparameters. In the first example, hyperparameter 1 is more important than hyperparameter 2. That means a small change in hyperparameter 1 influences the model performance more than a change in hyperparameter 2. As long as hyperparameter 1 is within the right values the model will be accurate, despite whatever value hyperparameter 2 takes. 

\section{Results and Discussion}

In this section, we examine the performance of both grid search and BOSS in tuning the hyperparameters of KRR-based machine learning models for predicting molecular HOMO energies based on molecular structures. An important objective is to establish whether BOSS is capable of finding similar hyperparameter solutions as the grid search algorithm, which is guaranteed to succeed. BOSS solutions of $\boldsymbol{\hat z}$, $f(\boldsymbol{\hat z})$ and $\mu(\boldsymbol{\hat z})$ are presented in Table \ref{table:results_qm9} alongside equivalent results from the grid search.

In this study we consider the case of CM and MBTR descriptors, which changes the dimensionality and complexity of the search. Concurrently, we analyze the effect of dataset type and training set size on the hyperparameter tuning procedure and optimal solutions. Finally, we compare timings of grid search and BOSS to estimate which approach is more efficient in each of the above described settings.

\subsection{KRR-CM hyperparameter tuning}

\begin{figure}[h!]
    \centerline{\includegraphics[width=\columnwidth]{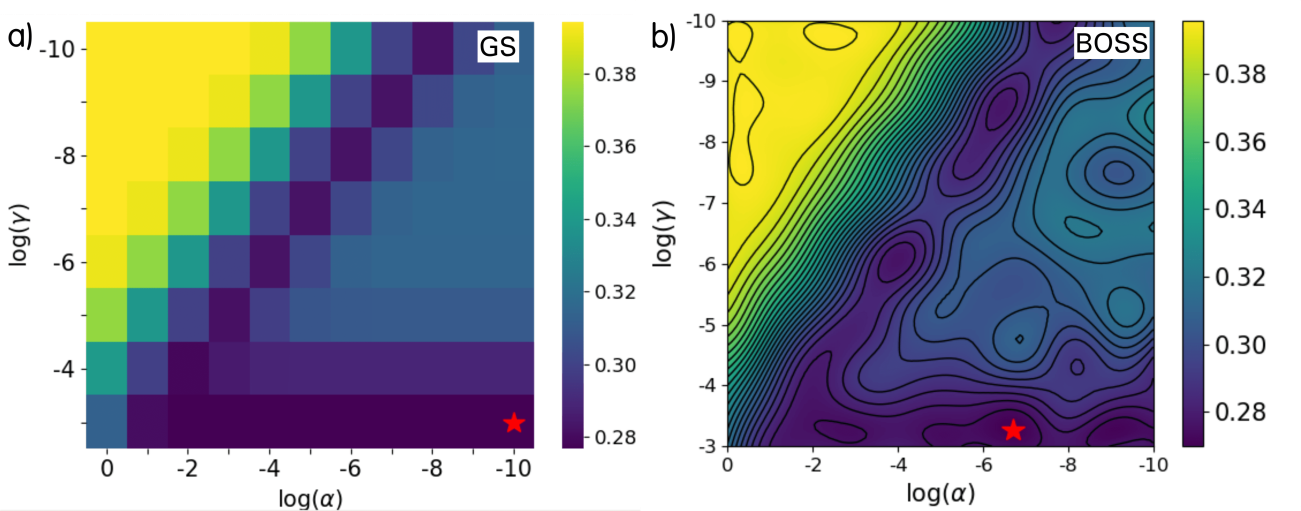}}
    \caption{MAE landscapes from the 2D optimization problem with the CM as molecular descriptor. a) Grid search MAE as a function of hyperparameters $\alpha$ and $\gamma$, evaluated on a logarithmic grid. b) BOSS model prediction $\mu(z)$ of the MAE as a function of log($\alpha$) and log($\gamma$). Both grid search and BOSS were applied to a subset of 2k molecules taken from the QM9 dataset. Optimal hyperparameters are shown as red stars.} 
\label{fig:qm9_cm}
\end{figure}

The CM materials descriptor has no parameters and the KRR kernel choice is clear. The MAE is thus a 2-dimensional function of KRR hyperparameters: regularization strength $\alpha$ and kernel width $\gamma$. Figure \ref{fig:qm9_cm} shows the 2-dimensional landscapes of MAE of a CM-KRR model for the QM9 dataset and a training set size of 2k. 
%Varied are the regularization strength $\alpha$ (x-axis) and kernel width $\gamma$ (y-axis). 
%Both are shown on a logarithmic grid. 
Panel a) depicts the grid search and panel b) the BOSS results, with logarithmic axes for clarity. All BOSS searches in this and subsquent sections are converged with respect to the number of acquisitions. The detailed convergence analysis will be presented in Section~\ref{sec:conv_timing}. 

It is clear that BOSS and grid search produce qualitatively similar MAE landscapes. The grid search landscape is naturally "pixelated", because it only has a 10$\times$10 resolution in Fig.~\ref{fig:qm9_cm}. Conversely, BOSS is not constrained to a grid and the Gaussian process in BOSS interpolates the MAE between the BOSS acquisitions. 
%Small irregularities in the BOSS-generated lanscape appear because the isotropic kernel ... maybe not needed.

Visualizing the MAE landscape tells us that the optimal parameter region has a complex and at first sight non-intuitive shape. The lowest MAE values in both methods lie on the diagonal of the hyperparameter landscape and along a horizontal line at the bottom of the landscape (for which $\gamma$ is $\sim$10$^{-3}$). Thus, the optimal parameter space has two parts: a co-dependent part, in which the choice of $\alpha$ and $\gamma$ is equally important for the KRR accuracy and a quasi one dimensional part, in which only the choice of $\gamma$ matters and $\alpha$ can assume any value. We will return to the analysis of this hyperparameter behavior in Section~\ref{sec:interpretation}.

Grid search and BOSS locate almost identical MAE minima: 0.277 eV for grid search and 0.270 eV for BOSS. The optimal hyperparameter solution $\boldsymbol{\hat z}$ found by BOSS and grid search are found in the same hyperparameter region of low log($\alpha$) values and high log($\gamma$) values. We conclude that BOSS is capable of reproducing grid search solutions to 1\% accuracy in this case. Since BOSS and grid search produce qualitatively and quantitatively similar results, we will only consider BOSS MAE landscapes in the remaining discussion.

%All values of $\boldsymbol{\hat z}$, $f(\boldsymbol{\hat z})$ and $g(\boldsymbol{\hat z})$ are shown in Table \ref{table:results_qm9}. 
%Note that the BOSS surrogate model gives a slightly higher  minimum (0.302~eV) than the actual MAE at the global minimum (0.300~eV) and slightly lower than the grid search value (0.304~eV). This could be due to the coursness of the grid search grid or the fact that the BOSS surrogate model has slight variations that would converge out in the limit of infinite data. However, with 0.002~eV this variation between the models is much below our required accuracy, which is at last one or two orders of magnitude higher.

\begin{figure}[h]
    \centerline{\includegraphics[width=\columnwidth]{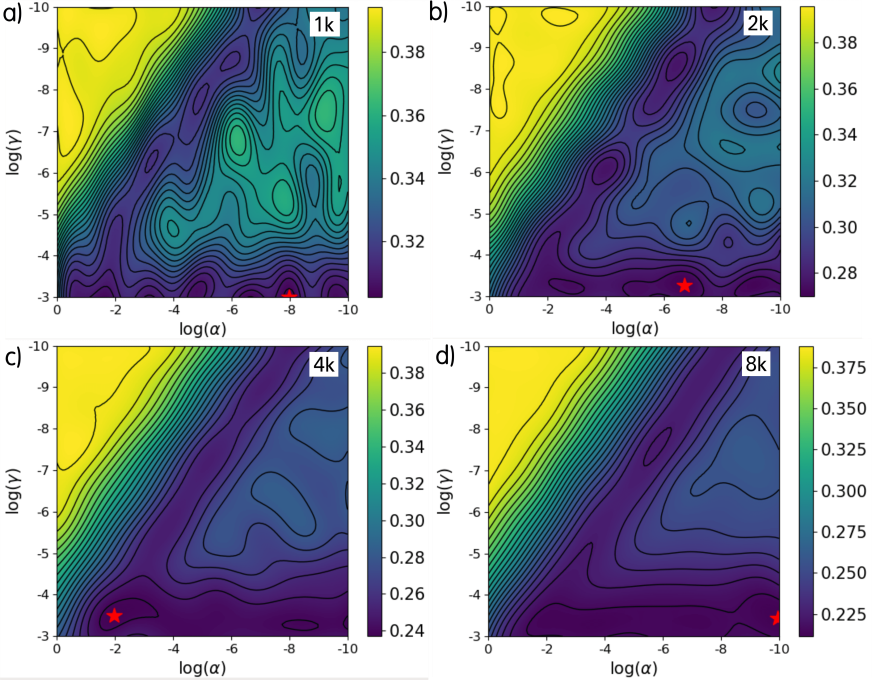}}
    \caption{BOSS KRR-CM hyperparameter landscapes for increasing dataset sizes of the QM9 dataset. The figure style is the same as in Fig.~\ref{fig:qm9_cm}.} 
\label{fig:qm9_trainsize}
\end{figure}

 Figure~\ref{fig:qm9_trainsize} illustrates BOSS MAE landscapes for increasing training set sizes of 1k, 2k, 4k and 8k molecules taken from the QM9 dataset. Tables~\ref{table:results} and \ref{table:results_qm9} summarize the optimal hyperparameter search. We find that the MAE landscapes become more homogeneous with increasing dataset size and that the optimal MAE is decreasing (as expected). The optimal solution of hyperparameters $\boldsymbol{\hat z}$ vary somewhat, but are always found within the horizontal region at the large $\gamma$ values. The slight variation is an indication of the flat MAE landscapes in the aforementioned triangular solution space, on which many hyperparameter combinations yield low MAE values.

\begin{figure*}[ht!]
    \centerline{\includegraphics[width=0.9\textwidth]{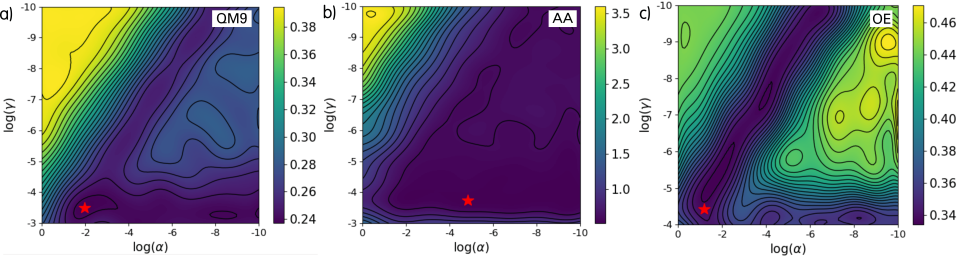}}
    \caption{MAE landscapes of 2D BOSS hyperparameter optimization with CM as molecular descriptor. Shown is the predicted MAE $\mu(x)$ as a function of hyperparameters $\alpha$ and $\gamma$, evaluated on a logarithmic grid for three different datasets a) QM9, b) AA and c) OE. From each dataset, a subset of 4k molecules was used. Optimal hyperparameters are shown as red stars.} 
\label{fig:cm_datasets}
\end{figure*}

Next, we compare the hyperparameter landscapes across the three different molecular datasets QM9, AA and OE, as shown in Fig.~\ref{fig:cm_datasets}. The model was trained on a subset of 4k molecules for each dataset. The dependence of MAE on the two KRR hyperparameters $\alpha$ and $\gamma$ is the same for all three datasets, with the the triangular region of optimal hyperparameters and flat MAE minima. However, the detailed dependence varies qualitatively with the dataset. For AA, the triangle is filled and we find a wide range of optimal hyperparameters in the landscape. This indicates that the KRR model is not overly sensitive to $\alpha$ and $\gamma$ for KRR-CM learning of the AA dataset. Conversely, for OE, the diagonal is more pronounced than the large $\gamma$ solution that dominates for QM9. The horizontal line of optimal hyperparameters has not yet developed for this training set size, indicating that broad feature widths $\gamma$ only lead to optimal learning when the regularization $\alpha$ is large. The optimal combination $\boldsymbol{\hat z}$ of hyperparameters differs across the three datasets (see Tables \ref{table:results} -\ref{table:results_oe} for optimal solutions $\boldsymbol{\hat z}$ and corresponding MAEs $f(\boldsymbol{\hat z})$ and $\mu(\boldsymbol{\hat z})$). Also the optimal MAE values vary considerably across the three datasets. This in accordance with our previous work, which revealed that the predictive power of KRR inherently depends on the complexity of the underlying dataset \cite{stuke_chemical_2019}. %\MT{the rest is redundant.} The lowest MAEs are achieved for QM9, which contains small molecules with similar and rather simple bonding patterns. The OE dataset includes larger molecules with complex structures and is more difficult for KRR to learn. MAEs are highest for the AA dataset of amino acids and dipeptides with different metal cations attached. 

%The optimal combination $\boldsymbol{\hat z}$ of hyperparameters differs across the three datasets (see Tables \ref{table:results} -\ref{table:results_oe} for optimal solutions $\boldsymbol{\hat z}$ and corresponding MAEs f$(\boldsymbol{\hat z})$ and g($\boldsymbol{\hat z}$)).

\subsection{KRR-MBTR 4D hyperparameter tuning}
%\subsection{4D hyperparameter tuning}

\begin{figure}[h]
    \centerline{\includegraphics[width=\columnwidth]{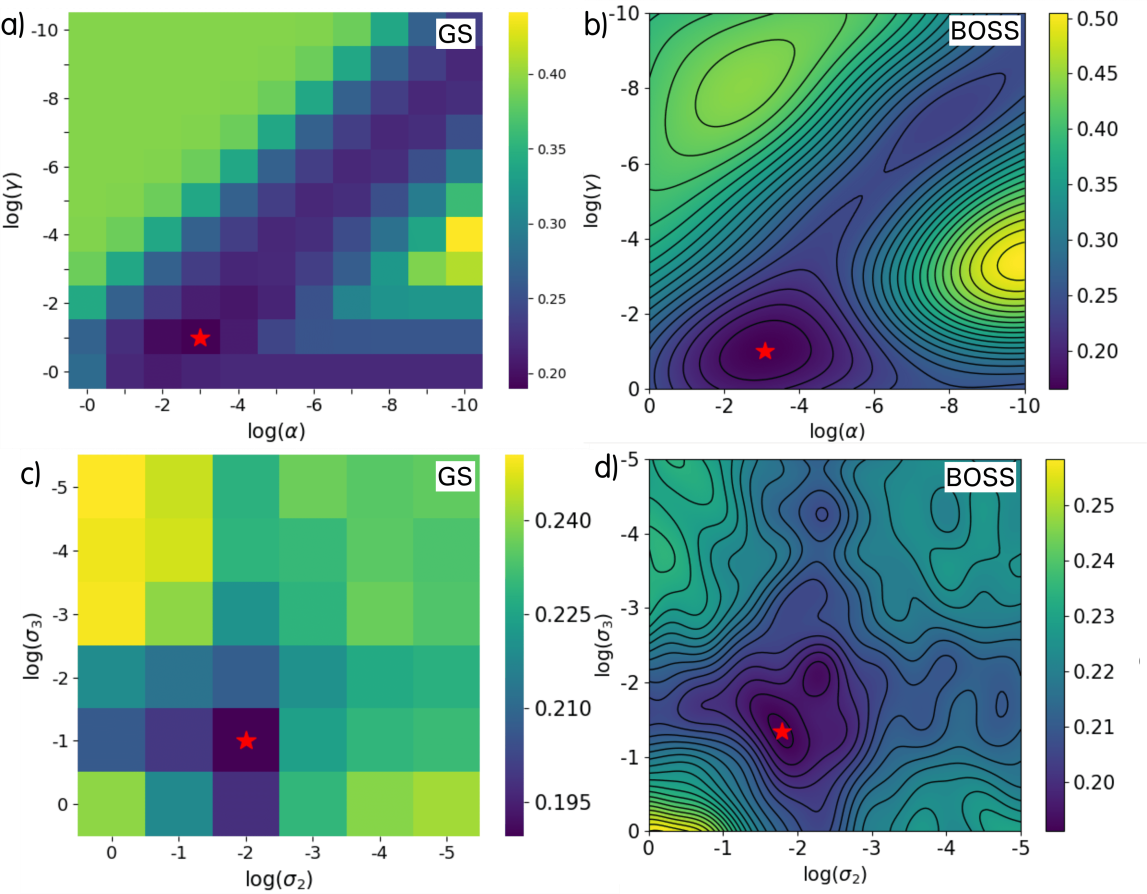}}
    \caption{MAE landscapes from the 4D hyperparameter optimization problem with the MBTR descriptor. Panels a) and b) show 2D slices through the logarithmic ($\alpha$, $\gamma$) plane, while panels c) and d) show 2D slices through the logarithmic ($\sigma_2$, $\sigma_3$) plane. In a) and c), the grid search MAE and in b) and d), the BOSS MAE surrogate model $\mu(x)$ are presented. Both grid search and BOSS were applied to a subset of 2k molecules taken from the QM9 dataset. Optimal hyperparameters are shown as red stars.} 
\label{fig:qm9_mbtr}
\end{figure}

We now consider the results of the 4D optimization problem, where the MBTR is used as molecular descriptor. %with fixed MBTR-scaling factors $s_2$ and $s_3$. 
BOSS builds a 4-dimensional surrogate model MAE($\alpha$,$\gamma$,$\sigma_2$, $\sigma_3$), which we compare to the reference 4-dimensional MAE landscape produced by grid search. Both 4D landscapes can be analyzed by considering 2-dimensional cross-sections. In Figure \ref{fig:qm9_mbtr}, we compare the BOSS surrogate model to the grid search result.

Figure \ref{fig:qm9_mbtr} a) and b) illustrate the (log($\alpha$), log($\gamma$)) cross-section of the four dimensional MAE landscapes, extracted at the global minimum $\boldsymbol{\hat z}$ with optimal $\sigma_2$, $\sigma_3$ values. Similar to the MAE landscapes of the 2D optimization problem, the optimal values lie on a diagonal and on a horizontal line at the bottom of the map. A notable difference to the previously discussed 2D case are the lower overall prediction errors. This is in line with our previous finding \cite{stuke_chemical_2019} that the MBTR encodes the atomic structure of a molecule better than the CM. 

In Figure \ref{fig:qm9_mbtr} c) and d), the 4D MAE landscapes are cut through the (log($\sigma_2$), log($\sigma_3$)) plane, while $\alpha$ and $\gamma$ are held constant at their optimal values. Here, the optimal MAEs are found only within a small region. In contrast to the KRR hyperparameters, the MAE is barely sensitive to $\sigma_2$ and $\sigma_3$, varying only by two decimals throughout the map (about 10\% of the value). All combinations of $\sigma_2$ and $\sigma_3$ are reasonably good choices for learning in this case. 

As in the two dimensional CM case, BOSS and grid search reveal qualitatively consistent hyperparameter landscapes and optimal solutions $\boldsymbol{\hat z}$, $f(\boldsymbol{\hat z})$ and $\mu(\boldsymbol{\hat z})$ (see Table \ref{table:results_qm9}). Therefore, we will only discuss MAE landscapes produced by BOSS in the following.

\begin{figure*}[tbh]
    \centerline{\includegraphics[width=0.8\textwidth]{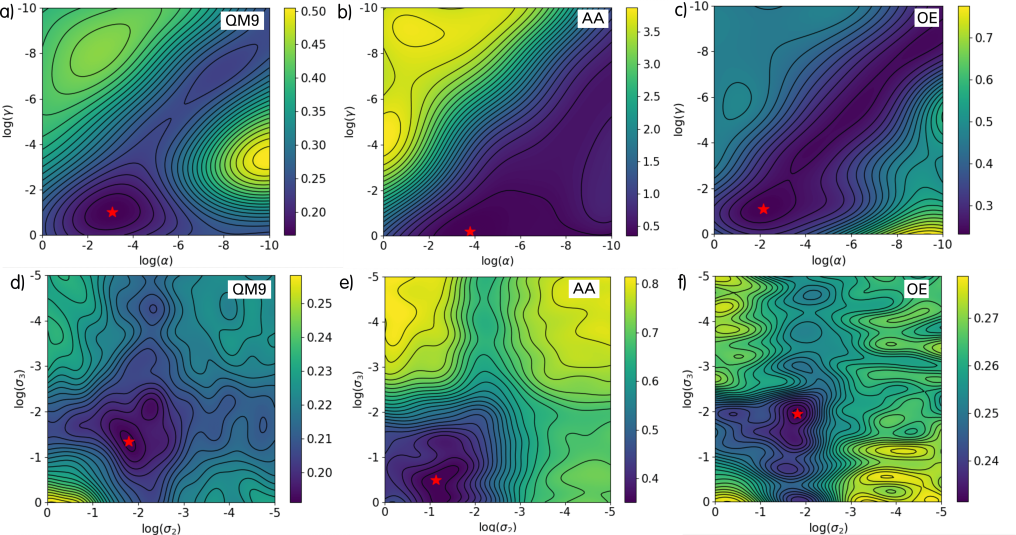}}
    \caption{2D-MAE landscapes from 4D BOSS hyperparameter optimization with MBTR as molecular descriptor. Shown is the predicted MAE $\mu(x)$ as a function of hyperparameters $\alpha$ and $\gamma$, evaluated on a logarithmic grid for three different datasets a) QM9, b) AA and c) OE. From each dataset, a subset of 2k molecules was used. Optimal hyperparameters are shown as red stars.} 
\label{fig:mbtr_datasets}
\end{figure*}

In Figure \ref{fig:mbtr_datasets}, we present BOSS MAE landscapes for the three different molecular datasets, where a subset of 2k molecules of each dataset was used for KRR. The comparison of the three MAE landscapes reveals that the QM9 dataset of small organic molecules is easiest to learn among all three datasets, since the MAE values are lowest. This is in accordance with the previous case of KRR-CM. We observe the highest MAEs for the AA dataset, which is most difficult to learn due to the higher chemical complexity of the molecules.

Panels a) - c) show MAE landscapes in logarithmic ($\alpha$ $\gamma$) planes and reveal the familiar diagonal pattern containing the lowest MAE. Compared to the 2D case with the CM descriptor, the landscapes are more homogeneous and the location $\boldsymbol{\hat z}$ of optimal hyperparameters lies within the same region for all three datasets. The choice of $\alpha$ and $\gamma$ seems to be independent of the dataset. Panels d) - f) show MAE landscapes in logarithmic ($\sigma_2$, $\sigma_3$) plane. All three datasets feature a cross-like shape of low MAE values. For QM9 and OE, the optimal MAE roughly lies on the crossover point. For these two datsets, the $\sigma_2$ and $\sigma_3$ values do not have a significant influence on the MAE range. For the AA dataset, in constrast, the choice of $\sigma_2$ and $\sigma_3$ dramatically affects the quality of learning and should be set correctly.

%\begin{figure}[h!]
%    \centerline{\includegraphics[width=\columnwidth]{weights.png}}
%    \caption{Scaling of weights for $k$=2 and $k$=3 terms for QM9 and OE.} 
%\label{fig:qm9_weights}
%\end{figure}

%\subsection{KRR-MBTR 6D hyperparameter tuning}
%\subsection{6D hyperparameter tuning}

\subsection{Interpretation of MAE landscapes}
\label{sec:interpretation}

\begin{figure}[h!]
    \centerline{\includegraphics[width=0.95\columnwidth]{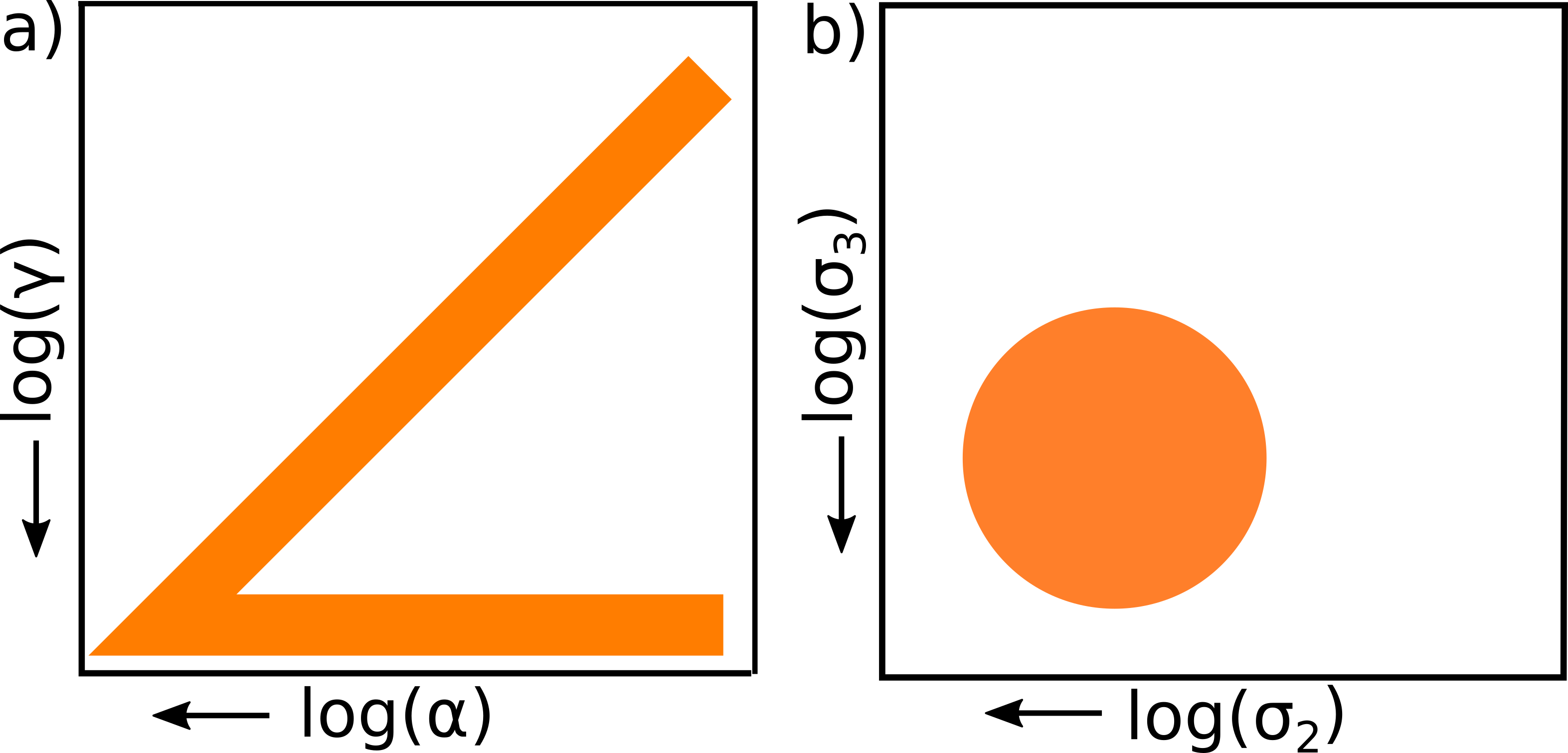}}
    \caption{Schematic depiction of the observed optimal hyperparameter regions for two different hyperparameter spaces. a) Typical landscape in ($\alpha, \gamma$)-plane. b) Typical landscape in ($\sigma_2, \sigma_3$)-plane.} 
\label{fig:landscapes}
\end{figure}

We now discuss the distinctively different optimal regions of different hyperparameter classes.  Figure~\ref{fig:landscapes} schematically depicts the shapes of these optimal hyper parameter regions for the KRR parameters ($\alpha$-$\gamma$ plane in panel a)) and the MBTR feature widths ($\sigma_2$-$\sigma_3$ plane in panel b)). % and the weighting factors ($s_2$-$s_3$ plane in panel c)). 

To understand the triangular shape in panel a), we have to recall the two kernels  \ref{eq:gaussian} and \ref{eq:laplacian} and the KRR regularization equation eq.~\ref{eq:krr_minimization}. For large feature widths $\gamma$, the abstract molecular space in which we measure distances between molecules $\boldmath{M}$ and $\boldmath{M'}$, is filled with broad Laplacians or Gaussians. The kernel expansion in eq.~\ref{eq:krr_minimization} then picks up a contribution from almost any molecular pair $\boldmath{M}$ and $\boldmath{M'}$. This means that the expansion coefficients have to be small. However, if the expansion coefficients are small, the regularization term is small and the size of the regularization strength $\alpha$ does not matter. This explains the horizontal line at the bottom of the triangle. 

For smaller values of the feature width $\gamma$, the Laplacians or Gaussians in molecular space become narrower, until, in the limit of an infinitely small $\gamma$, we obtain delta functions. The narrower the expansion functions are in molecular space, the larger the expansion coefficients need to be to give finite target properties. When the expansion coefficients increase in size, however, the regularization parameter needs to reduce to keep the size of the penalty term low. $\gamma$ and $\alpha$ become co-dependent, which explains the diagonal line in the triangle.

For the MBTR hyperparameters, the situation is qualitatively different. Although $\sigma_2$ and $\sigma_3$ are also associated with feature widths, they control the broadening of features in the structural representation of molecular bond distances and angles. For large broadenings (i.e. large $\sigma_2$ and $\sigma_3$), peaks associated with individual features in the MBTR might merge and the MBTR looses resolution. For very small broadenings (i.e. very small $\sigma_2$ and $\sigma_3$) features are represented by very narrow peaks, which may not be captured by the MBTR grids. The MBTR again looses resolution. The hyperparameter sweet spot therefore lies in a roughly circular region of  moderate $\sigma_2$ and $\sigma_3$ values. For our molecules and our datasets, the optimal $\sigma_2$ and $\sigma_3$ values are between 10$^{-1}$ and 10$^{-2}$, so closer to the bottom left corner of the hyperparameter landscape.

%$s_2$ and $s_3$ control the relative weighting of the bond distance and the angular term and, at the same time, limit the ranges of these terms in real-space. If both parameters are 0, bond distances and bond angles of all atom pairs in triples in the whole molecule are taken into account with equal weight regadless of how far apart the atoms are. This corresponds to maximum information content and concommittantly results in a low MAE. For molecules, much larger than the ones considered in the OE dataset, information overload might ensue and the MAE would go up, but we have not encountered such a situation, yet. 

%With increasing distance cutoff for the bonding term, bonding information is lost in the MBTR and the importance of the angular term grows. For our datasets, $s_3$ values of up to 1 or 2 do not yet seem to cutoff angular information, which explains the narrow rectangular optimal region at the bottom of the $s_2$-$s_3$ plane. For large $s_3$ values, angular information start getting lost in the MBTR and the MAE goes up. The situation is similar for $s_2$. If the range of the angular term is restricted by increasing $s_3$, the bond distance term has to compensate. Again, no information seems to be lost for $s_2$ values up to $\sim$2, which explains the rectangular region along the y-axis. For larger $s_2$, bond distance information is cut-off and the MAE increases. 

This section demonstrates that visualizing the hyperparameter landscapes greatly faciliates our understanding of the hyperparameter behavior in KRR and in machine learning in general. The BOSS methods provides an efficient way of generating easily readable landscapes that enable a deeper analysis of machine learning models.

\subsection{Convergence, scaling and computational cost}
\label{sec:conv_timing}

\subsubsection{Convergence}

\begin{figure}[htb]
    \centerline{\includegraphics[width=0.9\columnwidth]{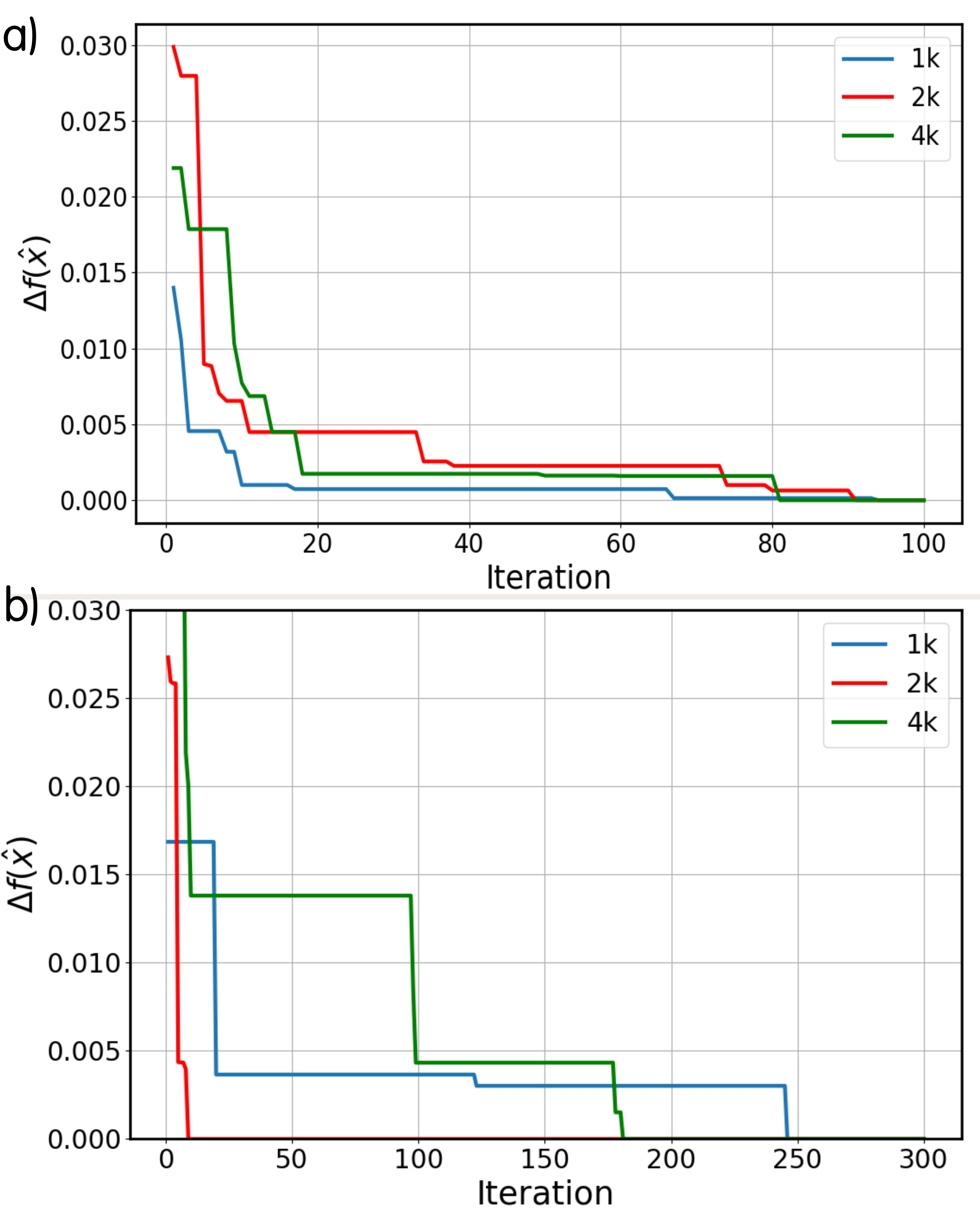}}
    \caption{BOSS convergence as a function of iterations, performed on QM9 for three different training set sizes. Panel a) shows a 2D search with the CM and panel b) a 4D search with the MBTR as molecular descriptor.  The convergence criterium $\Delta f(\hat x)$ describes the difference between the currently lowest $f(\hat x)$ and the lowest $f(\hat x)$ after the maximum number of iterations.} 
\label{fig:convergence}
\end{figure}

Figure~\ref{fig:convergence} illustrates the convergence of BOSS as a function of iterations, using the CM and the MBTR as molecular descriptor, for different training set sizes. Here, we consider the global minimum location $\boldsymbol{\hat z}$ in the landscape as the surrogate model improves, compute its true MAE value $f(\boldsymbol{\hat z})$ and track the lowest value observed. In the limit of the predefined maximum number of iterations (100 for 2D and 300 for 4D) the model no longer changes, so we adopt the final MAEs $f(\boldsymbol{\hat z})$ as zero reference. In the final step, we subtract the reference from the sequence of lowest MAE values observed to obtain the bare convergence $\Delta f(\boldsymbol{\hat z})$ of the MAE with BOSS iteration steps.
%Depicted on the $y$-axis is the logarithmic difference between the currently lowest value of the objective function evaluated at the model minimum prediction location $\boldsymbol{\hat z}$ and the lowest value of the objective function ever observed after reaching the predefined maximum number of iterations.  

Figure~\ref{fig:convergence} shows that $\Delta f(\boldsymbol{\hat z})$ drops quickly with BOSS iterations. Since the best MAE resolution we achieved with grid search was 0.02eV, we define the BOSS convergence criteria as $\Delta f(\boldsymbol{\hat z})$)$\leq 10^{-2}$. We find that in the 2D case (KRR-CM in Fig.~\ref{fig:convergence}a)), the BOSS solution is already converged in fewer than 20 iterations, regardless of training set size. In the 4D hyperparameter search (KRR-MBTR in Fig.~\ref{fig:convergence}b)), BOSS reaches convergence in less than 50 iterations with underlying datasets of sizes 1k and 2k. For a dataset size of 4k, it takes almost 100 iterations to reach convergence. Some variation in convergence behaviour is expected, and averages from repeated runs would provide better averaged results in future work. Nonetheless, the best hyperparameter solutions were clearly found within 100 iterations in all scenarios.

\subsubsection{Formal computational scaling}

Formally, the computational time of a KRR run for a fixed training set size can be estimated as follows
\begin{equation*}
    t_{\textnormal{total}} = n_{\textnormal{desc}} \cdot \widetilde{t}_{\textnormal{desc}} + n_{\textnormal{KRR}} \cdot \widetilde{t}_{\textnormal{KRR}} + t_{\textnormal{process}}.
\end{equation*}
$\widetilde{t}_{\textnormal{desc}}$ is the average time to build the molecular descriptor for all molecules and $n_{\textnormal{desc}}$ is the number of times the descriptor has to be generated. $\widetilde{t}_{\textnormal{KRR}}$ is the average time  to perform the 5-fold cross-validated KRR step to determine the regression coefficients $\boldmath{w}$ and $n_{\textnormal{KRR}}$ the number of times this has to be done. Finally, $\widetilde{t}_{\textnormal{process}}$ is extra time used by the BOSS method to refine the surrogate model and determine the location for the the next data point acquisition. 

$\widetilde{t}_{\textnormal{desc}}$ should scale linearly with training set size, since one descriptor per molecule has to be generated. Conversely, we expect $\widetilde{t}_{\textnormal{KRR}}$ to scale cubically with training set size, since the determination of the regression weights $\boldmath{w}$ in eq.~\ref{eq:solution} requires the inversion of the kernel matrix. The dimension of the kernel matrix grows linearly with training set size and its inversion will therefore scale cubically. $\widetilde{t}_{\textnormal{process}}$ only depends on the dimensionality of the search, but not on the training set size. Since $\widetilde{t}_{\textnormal{process}}$ is typically small, we will omit it from the timing discussion.

\subsubsection{Computational time}

%\PRc{We should add the 6D numbers to Table~\ref{table:mbtr_krr} and to the text.}

\begin{figure}[htb]
    \centerline{\includegraphics[width=1.05\columnwidth]{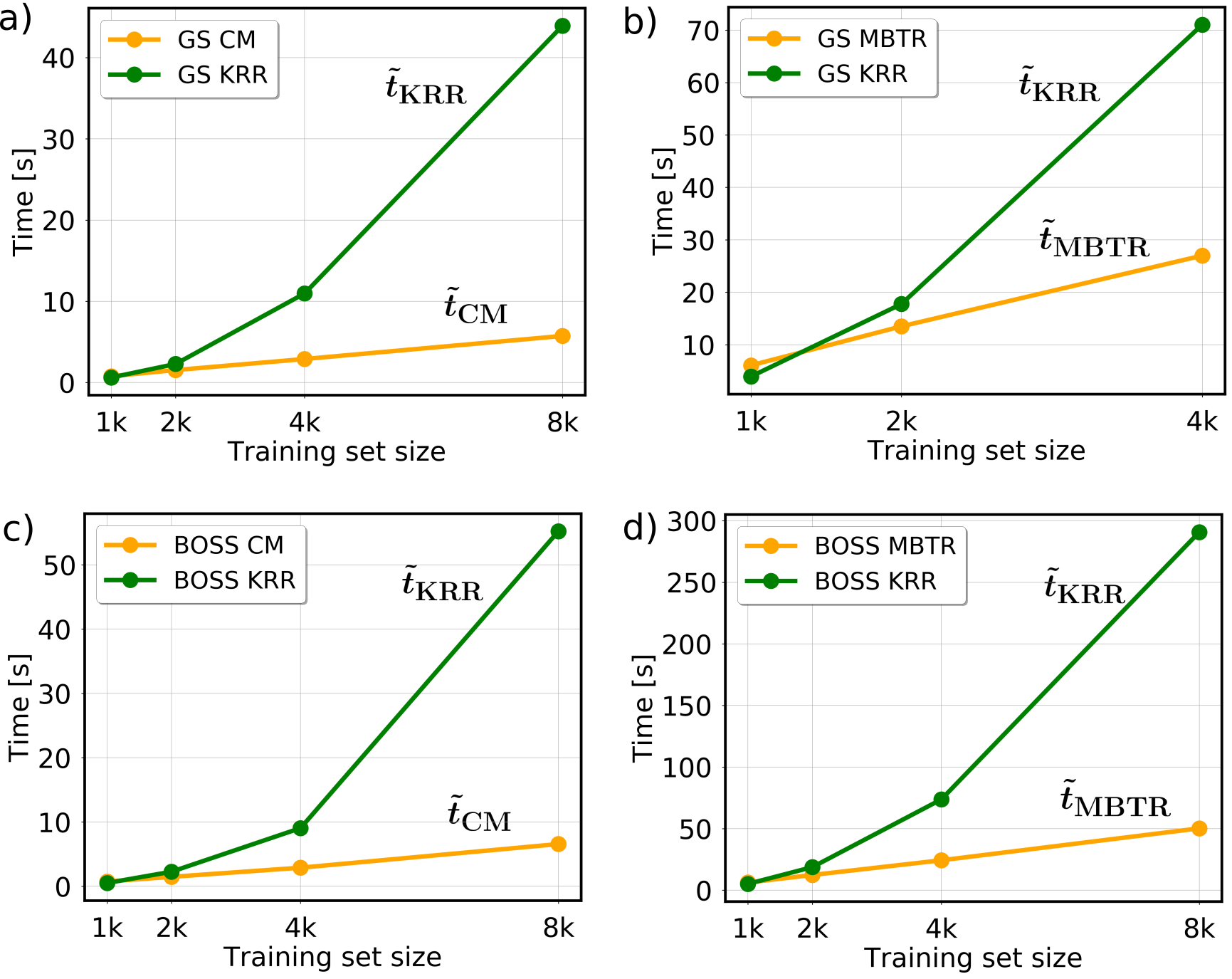}}
    \caption{Comparison between average times (per iteration) needed to perform 5-fold cross-validated KRR ($\widetilde{t}_{\textnormal{KRR}}$) and to build the molecular descriptor ($\widetilde{t}_{\textnormal{CM}}$ or $\widetilde{t}_{\textnormal{MBTR}}$). In a) and b), timings are shown for grid search (GS) and in c) and d) for BOSS. Panels a) and c) are for the CM (2D search) and b) and d) for the MBTR (4D search). Note that for the MBTR, grid search is only performed for training set sizes up to 4k. }
\label{fig:times}
\end{figure}

To collect run time information we added timing statements to our KRR implementation. Figure \ref{fig:times} depicts the average time the BOSS and grid search algorithms need to build the molecular descriptor and run cross-validated KRR, as a function of training set size for KRR-CM and KRR-MBTR. Panels a) and b) show the timings for grid search and panels c) and d) for BOSS. We observe that our formal scaling estimates in the previous section are confirmed, as the average time for building the CM ($\widetilde{t}_{\textnormal{CM}}$) or the MBTR ($\widetilde{t}_{\textnormal{MBTR}}$) grow linearly with training set size, while  $\widetilde{t}_{\textnormal{KRR}}$ grows cubically with training set size. Hence, for the smallest training set size of 1k, it might take less time to perform KRR than to build the molecular descriptor, while for training set sizes of 2k and larger the cubic scaling of the KRR part has already overtaken the descriptor building.

\begin{figure}[h!]
    \centerline{\includegraphics[width=0.9\columnwidth]{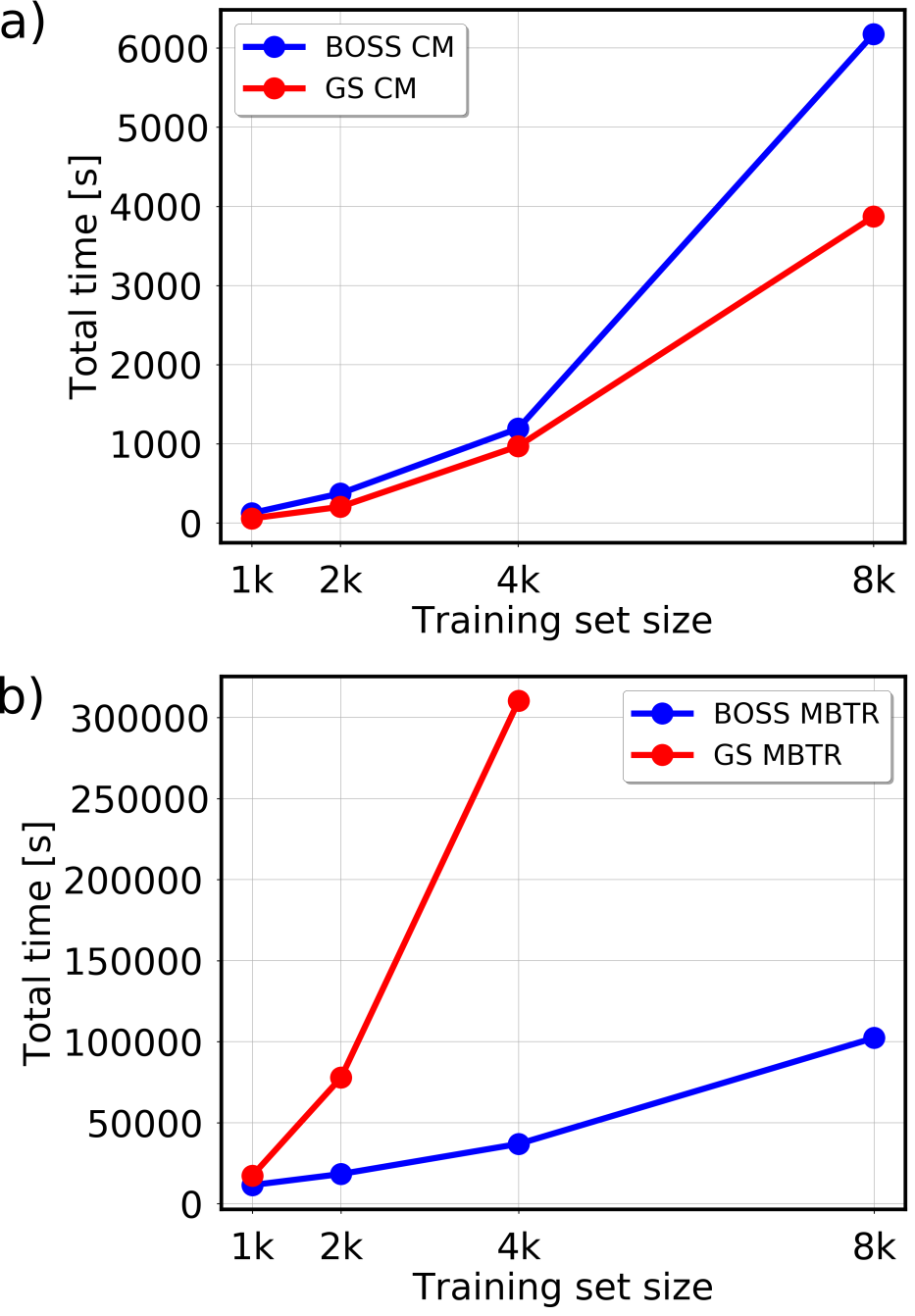}}
    \caption{Total times for hyperparameter optimization by BOSS and grid search as a function of training set size. In a) the CM and in b) the MBTR is used as molecular descriptor. Timings are shown for optimization on the QM9 dataset. For grid search (GS), only training set sizes up to 4k are feasable to be considered.}
\label{fig:total_times}
\end{figure}

The total computing time as a function of training set size is presented in Fig.~\ref{fig:total_times}. In the 2D case the grid search outperforms BOSS, while in the 4D case, BOSS is significantly faster than grid search. To determine which approach is faster -- grid search or BOSS -- it comes down to how often the descriptor has to be build, $n_{\textnormal{desc}}$, in each method and how often cross-validated KRR has to be performed, $n_\textnormal{KRR}$. Table~\ref{table:mbtr_krr} shows both numbers for grid search and BOSS. In grid search, $n_{\textnormal{desc}}$ and $n_{\textnormal{KRR}}$ are fixed numbers (see  Algorithms \ref{code_gs_cm} and \ref{code_gs_4d}). They depend only on the size of the hyperparameter grid. In KRR-CM, the CM needs to be computed only once at the beginning of the routine. Cross-validated KRR is performed 121 times, for each combination of $\alpha$ and $\gamma$. In a 4D KRR-MBTR grid search, the MBTR needs to be computed for each combination of $\sigma_2$ and $\sigma_3$, i.e. 36 times. Cross-validated KRR is performed for each possible combination of $\alpha$, $\gamma$, $\sigma_2$ and $\sigma_3$, i.e. 4,356 times. 

In BOSS, the molecular descriptor must be built and cross-validated KRR must be performed every single time the objective function is evaluated, as shown in Algorithms \ref{code_boss_2d} and \ref{code_boss_4d}. This means that $n_{\textnormal{desc}}$ and $n_{\textnormal{KRR}}$ solely depend on the number of BOSS iterations required to converge the MAE landscapes. 

Since $\widetilde{t}_{\textnormal{KRR}}$ scales cubically, the critical number is $n_{\textnormal{KRR}}$. As Tab.~\ref{table:mbtr_krr} illustrates, $n_{\textnormal{KRR}}$ are roughly the same for grid search and BOSS in the 2D search. Already for a 4D search, BOSS requires significantly fewer KRR evaluations than grid search and is computationally much more efficient.

\begin{table}
\centering
 \begin{tabular}{|c|c|c||c|c|} 
  \multicolumn{3}{@{}l}{}\\
 \hline
   &  \multicolumn{2}{c||}{\textbf{2D (CM)}} & \multicolumn{2}{c|}{\textbf{4D (MBTR)}} \\
   \hline
   \hline
 & GS & BOSS & GS & BOSS\\
 \hline
 $\boldsymbol{n}_{\textnormal{desc}}$ & 1  & 100& 36 & 300 \\ 
 \hline
 $\boldsymbol{n}_{\textnormal{KRR}}$ & 121  & 100 & 4,356 & 300 \\
 \hline
\end{tabular}
 \caption{Number of times the molecular descriptor is built ($n_{\textnormal{desc}}$) and cross-validated KRR is performed ($n_\textnormal{KRR}$), using the grid search (GS) and BOSS approaches.}
 \label{table:mbtr_krr}
\end{table}

\begin{table*}[ht!]
\centering
 \begin{tabular}{|c|c|c|c|c|c|c|c|c|c} 
 \multicolumn{2}{@{}l}{}\\
\hline
  &  & & \multicolumn{2}{c|}{\textbf{QM9}} & \multicolumn{2}{c|}{\textbf{AA}} & \multicolumn{2}{c|}{\textbf{OE}}\\
  \hline
 descriptor & hyperparam. & train size & BOSS  & GS & BOSS & GS & BOSS & GS \\
 \hline\hline
CM & $\alpha$, $\gamma$ & 1k & 0.300 [0.302] & 0.304 & 0.829 [0.812] & 0.843 & 0.355 [0.357] & 0.368 \\ 
CM & $\alpha$, $\gamma$ & 2k & 0.269 [0.270] & 0.277 & 0.645 [0.651] & 0.658 & 0.353 [0.350] & 0.355 \\ 
CM & $\alpha$, $\gamma$ & 4k & 0.237 [0.238] & 0.244 & 0.507 [0.508] & 0.509 & 0.332 [0.335] & 0.338 \\ 
CM & $\alpha$, $\gamma$ & 8k & 0.212 [0.211] & 0.215 & 0.373 [0.374] & 0.382 & 0.303 [0.305] & 0.309 \\
 %\hline
%MBTR & $\alpha$, $\gamma$ & 1k & 0.223 [0.225] & 0.231 & 0.485 [0.497] & 0.504 & 0.241 [0.242] & 0.254\\
%MBTR & $\alpha$, $\gamma$ & 2k & 0.204 [0.207] & 0.211 & 0.360 [0.365] & 0.349 & 0.229 [0.232] & 0.238 \\
%MBTR & $\alpha$, $\gamma$ & 4k & 0.180 [0.180] & 0.185 & 0.278 [0.280] & 0.281 & 0.212 [0.212] & 0.217\\
%\hline
%MBTR & $\sigma_2$, $\sigma_3$ & 1k & 0.211 [0.211] & 0.217 & 0.470 [0.482] & 0.512 & 0.239 [0.239] & 0.242 \\
%MBTR & $\sigma_2$, $\sigma_3$ & 2k & 0.189 [0.191] & 0.196 & 0.340 [0.351] & 0.367 & 0.228 [0.231] & 0.234 \\
%MBTR & $\sigma_2$, $\sigma_3$ & 4k & 0.171 [0.173] & 0.176 & 0.272 [0.274] & 0.297 & 0.215 [0.217] & 0.217 \\
\hline 
MBTR & $\alpha$, $\gamma$, $\sigma_2$, $\sigma_3$ & 1k & 0.207 [0.212] & 0.214 & 0.464 [0.466] & 0.500 & 0.246 [0.243] & 0.246 \\
MBTR & $\alpha$, $\gamma$, $\sigma_2$, $\sigma_3$ & 2k & 0.190 [0.165] & 0.190 & 0.338 [0.348] & 0.361 & 0.233 [0.231] & 0.227 \\
MBTR & $\alpha$, $\gamma$, $\sigma_2$, $\sigma_3$ & 4k & 0.159 [0.162] & 0.166 & 0.334 [0.276]& 0.296 & 0.216 [0.218] & 0.219 \\
MBTR & $\alpha$, $\gamma$, $\sigma_2$, $\sigma_3$ & 8k & 0.142 [0.149] & -- & 0.213 [0.214] & -- & 0.189 [0.184] & --\\
\hline
\end{tabular}
 \caption{MAEs [eV] for the optimal set of hyperparameter found by BOSS and grid search (GS). Results are depicted for three different molecular datasets QM9, AA and OE. For BOSS, the first value is the best ever observed true function value f($\boldsymbol{\hat z}$), evaluated at the predicted optimal point $\boldsymbol{\hat z}$. The second value in squared brackets is the global minimum predicted by the surrogate model, $\mu(\boldsymbol{\hat z})$, at maximum number of iterations. For GS, the depicted value corresponds to the best model performance $f(\boldsymbol{\hat z})$.}
 \label{table:results}
\end{table*}

%\section{Discussion}

\section{Conclusion}
In this work, we have used the Bayesian optimization tool BOSS to optimize hyperparameters in a KRR machine-learning model that predicts molecular orbital energies. We use two different molecular descriptors, the CM and the MBTR. While the CM has no hyperparameters, the MBTR molecular descriptor introduces %up to four hyperparameters 
two extra hyperparameters to the optimization problem. We therefore performed BOSS searches in spaces of up to four dimensions. We compared MAE landscapes in hyperparameter space and the efficiency of the BOSS approach with the commonly used grid search approach for three different molecular datasets.

For CM as molecular descriptor, only the two KRR hyperparameters $\alpha$ and $\gamma$ need to be optimized. The 2D landscapes in hyperparameter space produced by BOSS and grid search agree very well, with the lowest MAE values lying on a diagonal and a horizontal line. This is the case for all three datasets and for all training set sizes. 

For MBTR as molecular descriptor, %we first studied the 4D case where the KRR hyperparameters $\alpha$ and $\gamma$ and the two MBTR broadening widths $\sigma_2$ and $\sigma_3$ are optimized%, while fixing the MBTR scaling factors $s_2$ and $s_3$. 
MAE landscapes cut through the (log($\alpha$), log($\gamma$))-plane qualitatively correspond to the MAE landscapes of the 2D optimization problem, while the overall prediction errors are notably lower. In the ($\sigma_2$, $\sigma_3$)-plane, the optimal MAE values are confined to a small, roughly spherical region. The MAE is not very sensitive to $\sigma_2$ and $\sigma_3$, in contrast to $\alpha$ and $\gamma$. Hence, all combinations of $\sigma_2$ and $\sigma_3$ are reasonable choices for the MBTR.

In terms of efficiency, grid search outperforms BOSS in the 2D case, while in the 4D case, BOSS is significantly faster than grid search. This paves the way for high-dimensional hyperparameter optimization with Bayesian optimization.

%Grid search does not choose the next hyperparameters to evaluate based on previous results. It is completely uninformed by past evaluations, and as a result, often spends a significant amount of time evaluating uninteresting hyperparameters.

%The entire concept of Bayesian model-based optimization is to reduce the number of times the objective function needs to be run by choosing only the most promising set of hyperparameters to evaluate based on previous calls to the evaluation function. The next set of hyperparameters are selected based on a model of the objective function called a surrogate.

%By evaluating hyperparameters that appear more promising from past results, Bayesian methods can find better model settings than random search in fewer iterations. Bayesian model-based methods can find better hyperparameters in less time because they reason about the best set of hyperparameters to evaluate based on past trials.

%Bayesian optimization methods are known to be efficient because they choose the next hyperparameters in an informed manner. They spend a little more time selecting the next hyperparameters in order to make fewer calls to the objective function. The time spent selecting the next hyperparameters is inconsequential compared to the time spent in the objective function. 

\begin{acknowledgments}
We gratefully acknowledge the CSC-IT Center for Science, Finland, and the Aalto Science-IT project for generous computational resources. This study has received funding from the Magnus Ehrnrooth and the Finnish Cultural Foundation as well as the Academy of Finland (Project No. 316601). This article is based on work from COST Action 18234, supported by COST (European Cooperation in Science and Technology).
\end{acknowledgments}

\section*{References}
\bibliography{references}% Produces the bibliography via BibTeX.
\bibliographystyle{abbrv}

\appendix

\section{Tables}

\begin{table*}[ht!]
\centering
 \begin{tabular}{|c|c|cc|cc|cc|cc|cc|} 
 \multicolumn{2}{@{}l}{}\\
\hline
\textbf{QM9}  &  &  \multicolumn{2}{c|}{\textbf{log($\alpha$)}} & \multicolumn{2}{c|}{\textbf{log($\gamma$)}} & \multicolumn{2}{c|}{\textbf{log($\sigma_2$)}} & \multicolumn{2}{c|}{\textbf{log($\sigma_3$)}} & \textbf{$\mu(\boldsymbol{\hat z})$} & \textbf{$f(\boldsymbol{\hat z})$}\\
  \hline
 descriptor & train size & BOSS & GS & BOSS & GS & BOSS & GS & BOSS & GS & BOSS & GS \\
 \hline\hline
 CM   & 1k & -8.0  & -3.0 & -3.0 & -3.0 & -- & -- & -- & -- & 0.302 & 0.304\\
 CM   & 2k & -6.7  &-10.0 & -3.3 & -3.0 & -- & -- & -- & -- & 0.270 & 0.277\\
 CM   & 4k & -2.0  &-10.0 & -3.5 & -3.0 & -- & -- & -- & -- & 0.238 & 0.244\\
 CM   & 8k & -10.0 &-10.0 & -3.4 & -3.0 & -- & -- & -- & -- & 0.211 & 0.215\\
 \hline
 MBTR & 1k & -2.1 & -2.0 & -1.0 & -1.0 & -1.8 & -2.0 & -1.2 & -2.0 & 0.212 & 0.214\\
 MBTR & 2k & -3.1 & -3.0 & -1.0 & -1.0 & -1.1 & -2.0 & -1.9 & -1.0 & 0.165 & 0.190\\
 MBTR & 4k & -2.3 & -2.0 & -3.9 & -1.0 & -1.5 & -5.0 & -1.3 & -3.0 & 0.162 & 0.166\\
 MBTR & 8k & -1.6 &  --  &  0.0 &   -- & -1.6 & --   & -1.1 &   -- & 0.149 & --\\ 
\hline
\end{tabular}
 \caption{Optimal hyperparameters $\boldsymbol{\hat z}$ and corresponding best model performance $f(\boldsymbol{\hat z})$ for the QM9 dataset, computed by BOSS and grid search (GS).}
 \label{table:results_qm9}
\end{table*}

\begin{table*}[ht!]
\centering
 \begin{tabular}{|c|c|cc|cc|cc|cc|cc|} 
 \multicolumn{2}{@{}l}{}\\
\hline
\textbf{AA}  &  &  \multicolumn{2}{c|}{\textbf{log($\alpha$)}} & \multicolumn{2}{c|}{\textbf{log($\gamma$)}} & \multicolumn{2}{c|}{\textbf{log($\sigma_2$)}} & \multicolumn{2}{c|}{\textbf{log($\sigma_3$)}} & \textbf{$\mu(\boldsymbol{\hat z})$} & \textbf{$f(\boldsymbol{\hat z})$}\\
  \hline
 descriptor & train size & BOSS & GS & BOSS & GS & BOSS & GS & BOSS & GS & BOSS & GS\\
 \hline\hline
 CM   & 1k & -4.1 &-10.0 & -4.0 & -4.0 & -- & -- & -- & -- & 0.812 & 0.843\\
 CM   & 2k & -1.0 &-10.0 & -3.9 & -4.0 & -- & -- & -- & -- & 0.651 & 0.658\\
 CM   & 4k & -4.8 &-10.0 & -3.7 & -4.0 & -- & -- & -- & -- & 0.508 & 0.509\\
 CM   & 8k & -6.5 &-10.0 & -3.8 & -4.0 & -- & -- & -- & -- & 0.374 & 0.382\\
 \hline
 MBTR & 1k & -3.1 & -4.0 & -1.7 & 0.0 & -1.0 & -1.0 &  0.0 & 0.0 & 0.466 & 0.500 \\
 MBTR & 2k & -3.8 & -4.0 & -1.9 & 0.0 & -1.5 & -1.0 & -4.0 & 0.0 & 0.348 & 0.361\\ 
 MBTR & 4k & -9.2 & -5.0 & -2.6 & 0.0 & -1.1 & -1.0 & 0.0 & 0.0 & 0.276 & 0.296 \\
 MBTR & 8k & -3.3 &  --  & 0    & --  & -1.8 & --   & -0.3 & -- & 0.214 & -- \\
\hline
\end{tabular}
 \caption{Optimal hyperparameters $\boldsymbol{\hat z}$ and corresponding best model performance $f(\boldsymbol{\hat z})$ for the AA dataset, computed by BOSS and grid search (GS).}
 \label{table:results_aa}
\end{table*}

\begin{table*}[ht!]
\centering
 \begin{tabular}{|c|c|cc|cc|cc|cc|cc|} 
 \multicolumn{2}{@{}l}{}\\
\hline
\textbf{OE}  &  &  \multicolumn{2}{c|}{\textbf{log($\alpha$)}} & \multicolumn{2}{c|}{\textbf{log($\gamma$)}} & \multicolumn{2}{c|}{\textbf{log($\sigma_2$)}} & \multicolumn{2}{c|}{\textbf{log($\sigma_3$)}} & \textbf{$\mu(\boldsymbol{\hat z})$} & \textbf{$f(\boldsymbol{\hat z})$}\\
  \hline
 descriptor & train size & BOSS & GS & BOSS & GS & BOSS & GS & BOSS & GS & BOSS & GS \\
 \hline\hline
 CM   & 1k & -3.0 & -6.0 & -6.7 & -10.0 & -- & -- & -- & -- & 0.357 & 0.368 \\
 CM   & 2k & -1.6 & -5.0 & -4.7 & -9.0  & -- & -- & -- & -- & 0.350 & 0.355\\
 CM   & 4k & -1.2 & -1.0 & -4.4 & -5.0  & -- & -- & -- & -- & 0.335 & 0.338 \\
 CM   & 8k & -1.4 & -6.0 & -4.4 & -10.0 & -- & -- & -- & -- & 0.305 & 0.309 \\
 \hline
 MBTR & 1k & -1.8 & -2.0 & -1.0 & -1.0 & -2.6 & -2.0 & -5.0 & -1.0 & 0.243 & 0.246\\
 MBTR & 2k & -2.2 & -2.0 & -1.1 & -1.0 & -2.6 & -2.0 & -1.1 & -2.0 & 0.231 & 0.227 \\
 MBTR & 4k & -1.8 & -2.0 & -1.0 & -1.0 & -2.8 & -3.0 & -1.2 & -2.0 & 0.218 & 0.219\\
 MBTR & 8k & -2.6 &  --  & -1.3 &  --  & -5.0 &  --  & -3.6 &  --  & 0.184 & -- \\
\hline
\end{tabular}
 \caption{Optimal hyperparameters $\boldsymbol{\hat z}$ and corresponding best model performance $f(\boldsymbol{\hat z})$ for the OE dataset, computed by BOSS and grid search (GS).}
 \label{table:results_oe}
\end{table*}

\end{document}